\newif \ifacs
\acsfalse

\ifacs
    \documentclass[journal=jpcafh,manuscript=article,super]{achemso}
    \setkeys{acs}{articletitle = true}
    \usepackage[titletoc,toc,title]{appendix}
    \SectionNumbersOn
\else
    \documentclass[aip,jcp,amsmath,amssymb,reprint]{revtex4-2}
\fi

\usepackage[a-1b]{pdfx} 
\usepackage{graphicx}
\usepackage{dcolumn}
\usepackage{bm}
\usepackage[utf8]{inputenc}
\usepackage[T1]{fontenc}
\usepackage{listings}
\usepackage{rotating}
\usepackage{color}
\usepackage{dcolumn} 
\usepackage{multirow} 
\usepackage{pifont} 
\usepackage{epsfig}
\usepackage{amsmath} 
\usepackage{subfigure}
\usepackage{float}
\usepackage{booktabs}
\usepackage{tabularx}
\usepackage{natbib}
\usepackage{gensymb}
\usepackage{rotating}
\usepackage{threeparttable}
\usepackage{comment}
\usepackage[normalem]{ulem}
\usepackage{physics}
\usepackage{mathrsfs}

\ifacs
\else
    
\begin{document}
\fi
  
\newcommand{\braz}{\langle 0^\text{e}|\langle 0^\text{p} |}
\newcommand{\ketz}{| 0^\text{e}\rangle| 0^\text{p} \rangle}
\newcommand{\Hpf}{\hat{H}_{\rm CS}}

\author{Stephen H. Yuwono}
\affiliation{
             Department of Chemistry and Biochemistry,
             Florida State University,
             Tallahassee, FL 32306-4390}   
\author{Katie A. Crouch}
\affiliation{
             Department of Chemistry,
             University of Houston,
             Houston, TX 77204-5003}  
\author{A. Eugene DePrince III}
\email{adeprince@fsu.edu}
\affiliation{
             Department of Chemistry and Biochemistry,
             Florida State University,
             Tallahassee, FL 32306-4390}

\title{Dynamic Response Functions for Cavity Quantum Electrodynamics Hartree--Fock Theory}

\ifacs
    \begin{document}
\else
\fi

\begin{abstract}

Frequency-dependent linear and quadratic response functions are implemented for a cavity quantum electrodynamics (QED) generalization of Hartree--Fock (HF) theory. Dynamic electric dipole polarizability, optical rectification hyperpolarizability, and second harmonic generation hyperpolarizability tensors are evaluated for molecules  strongly coupled to a single-mode optical cavity, and the results are benchmarked against the same tensors obtained from real-time time-dependent QED-HF simulations. Substantial cavity-induced changes to the frequency-dependent properties are observed for certain perturbing frequencies at large electron--photon coupling strengths. We also find special perturbing frequencies at which there is no cavity effect, regardless of the coupling strength. 
\end{abstract}

\maketitle

\section{Introduction}
\label{sec:introduction}

Light--matter interactions are a fundamental and ever-present theme in chemistry, physics, and materials science, due to their central role in a variety of contexts, including photocatalysis,\cite{Yoon14_1239176,Ng20_030903} sensing,\cite{Cappellaro17_035002,Lloyd18_724,Duan21_2100049,Nam24_061001} material property manipulation,\cite{Qiu17_e17039,Ishihara21_1885991,Liu22_031303,Jiang24_2301105} and coherent control of reaction dynamics.\cite{Dantus01_639,Ohmori09_487,Silberberg09_277,Amitay15_233003} In particular, the last decade has seen a steadily rising interest in the role that polariton states, which are created by strong coupling between photon and matter degrees of freedom, can play in such applications.\cite{Ebbesen21_eabd0336,Xiong23_776,Xiong24_2512} These hybrid light--matter states have shown promise in diverse areas spanning energy\cite{Lidzey14_712,Sanvitto17_e16212,Schwartz18_105,Forrest20_2002127,Lidzey21_16661,GomezRivas22_123,Musser22_2105569,Schwartz23_338} or charge\cite{Giebink20_177401,Ebbesen20_10219,George21_13616} transport phenomena, chemistry and catalysis,\cite{Ebbesen12_1592,Ebbesen16_11462,Shegai18_eaas9552,Ebbesen19_615,Ebbesen19_15324,George19_10635,Uji-i20_5332,Ebbesen20_10436,Borjesson21_2010737,Ebbesen21_16877,Shalabney21_chemrxiv.7234721.v5,George21_379,Ebbesen21_5712,George22_195,Jeffrey22_429,Nitzan22_43} and quantum science applications.\cite{Schoelkopf04_162,Mooij10_237001,Semba17_44,Nori19_19,Solano19_025005} 

Given the breadth and potential impact of the experimental studies on strong light--matter coupling, it is not surprising that the computational science community has embarked on substantial endeavors to develop both effective theories\cite{YuenZhou18_6325} for modeling collective strong coupling effects and cavity quantum electrodynamics (QED) generalizations of familiar electronic structure methods\cite{DePrince23_041301,Zhang25_10035} for predictive simulations in the single-molecule strong-coupling limit. Here, we focus on the latter efforts, within the domain of quantum chemistry. Most of the electronic structure portfolio has been extended to treat photons quantum mechanically within the cavity QED framework, from mean-field approaches (Hartree--Fock [HF], density functional theory [DFT], and time-dependent [TD] DFT),\cite{Bauer11_042107,Rubio14_012508,Rubio15_093001,Rubio18_992,Appel19_225,Rubio19_2757,Narang20_094116,Rubio22_7817,Rubio23_11191,Narang23_383,DePrince22_9303,Rubio22_094101,DePrince23_5264,DePrince24_064109,Tokatly13_233001,Rubio17_3026,Tokatly18_235123,Varga22_194106,Shao21_064107,Shao22_124104,Foley22_154103,Wilson24_094111} to correlated approaches utilizing a single-reference\cite{Narang23_arXiv:2307.14822, Reichman24_1143,Koch20_041043, Manby20_023262, Corni21_6664, DePrince21_094112, Flick21_9100, Koch21_094113, Koch22_234103, Flick22_4995, Rubio22_094101, Knowles22_204119, DePrince22_054105, Rubio23_2766, Rubio23_10184, Koch23_4938, Koch23_8988, Koch23_031002, DePrince23_5264,DePrince24_064109,Koch24_8876,Stopkowicz24_9572,Koch24_8838,Koch24_e1684,Koch25_021040,Dreuw23_124128,Koch25_027901,Koch25_arXiv:2510.19558} or multi-reference\cite{DePrince22_053710,Yu24_032804,Foley24_1214,Foley25_chemrxiv-2025-q6rfm,Ronca25_6862,Huo23_2353,Huo24_16184,Foley24_174105} frameworks. The majority of these studies have emphasized simulating energetics or dynamics of cavity-bound systems, while far fewer have examined cavity-induced changes to response properties,\cite{Koch24_e1684,Koch24_7841,Koch25_4447,Narang23_383,Narang24_369,DePrince25_8024} which are focus of this contribution. 

In this work, we extend our previous implementation\cite{DePrince25_8024} of QED-HF response theory for static molecular response properties to evaluate frequency-dependent linear and quadratic response functions. In particular, we focus on dynamic electric dipole polarizability and first hyperpolarizability tensors for cavity-bound molecular systems. In order to verify the correctness of our simulated response functions, we have also implemented an equivalent response theory formalism based on the explicit time evolution of the QED-HF wave function, which is based on the real-time (RT) TD finite-difference approach outlined in Refs.~\citenum{Li13_064104,Pedersen23_154102,Crawford25_1908}. We then apply the numerically verified frequency-dependent QED-HF response theory code to several molecular systems (formaldehyde, urea, and $p$-nitroaniline [pNA]) to investigate the extent of cavity-induced changes to the linear and quadratic response functions.

The remainder of this paper is structured as follows. In Section \ref{sec:theory}, we { provide} an overview of cavity QED-HF theory, followed by detailed derivations of the working equations for both dynamic response theory and RT-TD simulations at the cavity QED-HF level. Section \ref{sec:comp_details} provides the details of our calculations, the results of which are then reported in Section \ref{sec:results_discussion}. We provide brief concluding remarks in Section \ref{sec:conclusions}.

\section{Theory}
\label{sec:theory}

We begin by introducing the Pauli--Fierz (PF) Hamiltonian, represented in the length gauge and within the dipole and Born--Oppenheimer approximations. For interactions between the molecular degrees of freedom and a single optical cavity mode, we have
\begin{align}
        \label{EQN:PFH}
\hat{H}_\text{PF} = {}
    & \hat{H}_\text{e} + \omega_\text{cav} \hat{b}^\dagger\hat{b} - \sqrt{\frac{\omega_\text{cav}}{2}} {\bm \lambda}\cdot \hat{\bm{\mu}} \left(\hat{b}^\dagger + \hat{b}\right) \nonumber \\
    &{} + \frac{1}{2} \left({\bm \lambda}\cdot \hat{\bm{\mu}}\right)^2
\end{align}
The first two terms in this expression correspond to the isolated many-electron  and photon Hamiltonians, respectively. In the isolated photon Hamiltonian, the symbols $\hat{b}^\dagger$ and $\hat{b}$ are photon creation and annihilation operators, respectively, and $\omega_\text{cav}$ represents the frequency of the cavity mode. The molecular and photon degrees of freedom interact via the bilinear coupling {(BLC)} term (the third term in Eq.~\ref{EQN:PFH}), in which the coupling vector, ${\bm \lambda}$, and molecular dipole operator, $\hat{\bm \mu}=\hat{\bm \mu}^\text{e} + \hat{\bm \mu}^\text{nu}$ appear. Here, the superscripts ``e'' and ``nu'' refer to the electronic and nuclear components of the dipole operator, respectively. Within the present model, the magnitude of the coupling vector is related to an effective mode volume, $V_\text{eff}$, as
\begin{align}
    \lambda = |{\bm \lambda}| = \sqrt{\frac{4\pi}{V_\text{eff}}}
\end{align}
The last term in Eq.~\ref{EQN:PFH} is the dipole self-energy (DSE), which arises from the {length} gauge transformation applied to the minimal coupling, or "${\rm p} \cdot {\rm A}$" Hamiltonian,\cite{Huo20_9215} originally represented within the Coulomb gauge. We refer to Refs.~\citenum{Huo20_9215,Huo23_9786, DePrince23_041301} for additional details regarding the derivation of the PF Hamiltonian.

{The response of the coupled light-matter system to a time-dependent external perturbation is governed by the Hamiltonian
\begin{align}
\label{EQN:TD_HAMILTONIAN}
    \hat{H}(t) = \hat{H}_\text{PF} + V(t)
\end{align} 
Here, we take $V(t)$ to be a periodic time-dependent perturbation, with period $T$, that can be expressed as
\begin{align}
\label{EQN:PERTURBATION_T}
    {V(t) = \sum_{\omega} e^{-i\omega t}\, V(\omega)}
\end{align}
where the sum runs over a discrete set of frequencies closed under $\omega \to -\omega$.
The Fourier components of the perturbation, $V(\omega)$, can be expanded as
\begin{align}
\label{EQN:PERTURBATION_W}
    V(\omega) = \sum_A\epsilon_A(\omega)\hat{\Omega}_A
\end{align}
where $\hat{\Omega}_A$ is an operator representing the perturbation, and $\epsilon_A(\omega)$ is the corresponding strength at frequency $\omega$. We choose $\epsilon_A(-\omega)=\epsilon_A(\omega)^{*}$ so that $V(t)$ is Hermitian.
In this work, we focus on a perturbing electric field, in which case $V(\omega) = -\sum_A \epsilon_A(\omega) \hat{\mu}^{\text{e}}_A$.
} 

{ 
We note that, while $\hat{H}_\text{PF}$ includes a DSE contribution associated with the
cavity mode, we include no such term for the classical external field. Both cases can
be understood within the same Pauli--Fierz framework that led to 
Eq.~\ref{EQN:PFH},\cite{Rubio18_034005} but they differ in regime in which the coupling vector applies. The cavity mode has a small, fixed effective volume, so $\bm \lambda$ is finite. On the other hand, the external perturbing field is effectively unconfined, with
$V_\text{eff} \to \infty$ and $\bm \lambda \to 0$. This unconfined mode is occupied by a correspondingly
large number of photons, $N \to \infty$, such that the physical field strength (which is 
proportional to $\bm \lambda \sqrt{N}$) remains finite. In this limit, the BLC term reduces to the familiar semiclassical
interaction, $-\hat{\bm \mu}^\text{e} \cdot \bm \epsilon(t)$ (with
$\bm \epsilon(t) \equiv \sum_\omega \bm \epsilon(\omega) e^{-i\omega t}$). The DSE term, however, is vanishingly small in this limit, as it includes no compensating factor of $N$. Similarly, direct cross coupling terms between the cavity and external
field would also arise in the DSE, but these, too, vanish in this limit. 
}

\subsection{QED-HF Theory}

The QED-HF wave function is a direct product of a Slater determinant of electronic spin-orbitals, $|0^\text{e}\rangle$, and a zero-photon state, which we represent as the action of the coherent-state (CS) transformation operator, $\hat{U}_\text{CS}$, on the photon vacuum state, $|0^\text{p}\rangle$:
\begin{equation}
\label{EQN:QED_HF}
    |\Phi_{0}\rangle = |0^\text{e}\rangle \hat{U}_\text{CS}|0^{\rm p}\rangle
\end{equation}
The CS transformation operator can be defined analytically by \cite{Koch20_041043}
\begin{equation}
\label{EQN:U_CS}
    \hat{U}_{\rm CS} = {\rm exp}\left( \frac{-{\bm \lambda} \cdot \langle {\hat{\bm{\mu}}} \rangle }{\sqrt{2 \omega_{\rm cav}}} \left (\hat{b}^{\dagger} - \hat{b} \right ) \right) 
\end{equation} 
where the expectation value is taken with respect to $|0^\text{e}\rangle$. Taking the expectation value of $\hat{H}_\text{PF}$ with respect to $| \Phi_0\rangle$ gives us the QED-HF energy
\begin{align}
    E_0 &= \braz \hat{U}_\text{CS}^\dagger \hat{H}_\text{PF} \hat{U}_\text{CS}\ketz \\
    &= \braz \hat{H}_\text{CS}\ketz \\
    \label{EQN:QED_HF_ENERGY} 
    &= \langle 0^\text{e}| \hat{H}_\text{e} |0^\text{e}\rangle + \frac{1}{2} \langle 0^\text{e} | \left({\bm \lambda}\cdot[\hat{\bm{\mu}}^\text{e} - \langle \hat{\bm{\mu}}^\text{e}\rangle ]\right)^2 |0^\text{e}\rangle
\end{align}
where, on the third line, we have integrated out the photon degrees of freedom, and we can see that  $|0^\text{e}\rangle$ and $E_0$ can be determined from a HF procedure involving a dressed electronic Hamiltonian. We have also introduced the symbol $\hat{H}_\text{CS}$, which refers to the CS-transformed Hamiltonian and has the form
\begin{align}
\hat{H}_\text{CS} = {}& \hat{U}_\text{CS}^\dagger \hat{H}_\text{PF} \hat{U}_\text{CS} \nonumber \\
        \label{EQN:PFH_COHERENT}
      {} = {} & \hat{H}_\text{e} + \omega_\text{cav} \hat{b}^\dagger\hat{b} - \sqrt{\frac{\omega_\text{cav}}{2}} {\bm \lambda}\cdot [\hat{\bm{\mu}}^\text{e} - \langle \hat{\bm{\mu}}^\text{e}\rangle ] \left(\hat{b}^\dagger + \hat{b}\right) \nonumber \\
      & {} + \frac{1}{2} \left({\bm \lambda}\cdot [\hat{\bm{\mu}}^\text{e} - \langle \hat{\bm{\mu}}^\text{e}\rangle ]\right)^2
\end{align}
Note that, in obtaining expressions for $E_0$ and $\hat{H}_\text{CS}$, the nuclear contribution to the dipole vanishes because the expectation value of $\hat{\bm \mu}^\text{nu}-\expval{\hat{\bm \mu}^\text{nu}}$ is evaluated with fixed nuclei.

\subsection{Dynamic Response Theory for QED-HF}
\label{SEC:RESPONSE_THEORY}

In this section, we derive the working equations for QED-HF response theory by applying the formalism of Ref.~\citenum{Hattig98_1} to the QED-HF problem. An alternative derivation can be found in Ref.~\citenum{Koch24_e1684}. The TD response of the QED-HF wavefunction to a periodic external perturbation can be parametrized with an exponential unitary operator acting on the QED-HF reference state
\begin{align}
\label{EQN:TD_WFN}
    |\Phi(t)\rangle = e^{-i \phi(t)} e^{\hat{\kappa}(t)} |\Phi_0\rangle
\end{align}
Here, $\phi(t)$ is a phase factor, and $\hat{\kappa}(t) = \hat{\kappa}^\text{e}(t) + \hat{\kappa}^\text{p}(t)$ is an anti-Hermitian operator that encodes the electronic and photonic responses of the system to the applied perturbation, with the electronic and photonic parts defined by 
\begin{align}
\label{EQN:KAPPA_E}
    \hat{\kappa}^\text{e}(t) &= \sum_{ia} \left ( X^\text{e}_{ia}(t) \hat{a}^\dagger_a \hat{a}_i - Y^\text{e}_{ia}(t) \hat{a}^\dagger_i \hat{a}_a \right )\\
\label{EQN:KAPPA_P}
    \hat{\kappa}^\text{p}(t) &= X^\text{p}(t) \hat{b}^\dagger - Y^\text{p}(t) \hat{b}
\end{align}
In Eq.~\ref{EQN:KAPPA_E}, we have introduced the Fermionic creation ($\hat{a}^\dagger_a$, $\hat{a}^\dagger_i$) and annihilation operators ($\hat{a}_a$, $\hat{a}_i$) for electronic spin-orbital $\phi_a$ and $\phi_i$; throughout this work, the labels $a$, $b$, $c$, etc. and $i$, $j$, $k$, etc. will refer to electronic spin-orbitals that are unoccupied or occupied in $|0^\text{e}\rangle$, respectively. Because $\hat{\kappa}(t)$ is anti-Hermitian, we require $X^\text{e}_{ia}(t) = [Y^\text{e}_{ia}(t)]^*$ and $X^\text{p}(t) = [Y^\text{p}(t)]^*$. Expanding $|\Phi_0\rangle$, we could also write Eq.~\ref{EQN:TD_WFN} as
\begin{align}
    |\Phi(t)\rangle &= e^{-i \phi(t)} e^{\hat{\kappa}(t)} \hat{U}_\text{CS} \ketz \\
    & = \hat{U}_\text{CS} e^{-i\phi(t)}e^{i\theta(t)}e^{\hat{\kappa}(t)}\ketz
\end{align}
where $\theta(t)$ is a new phase factor arising from the fact that $\hat{U}_\text{CS}$ and $e^{\hat{\kappa}(t)}$ do not commute. As noted in Ref.~\citenum{Koch24_e1684}, this factor can simply be absorbed into $\phi(t)$, so we re-define the TD state as
\begin{align}
\label{EQN:TD_WFN_FINAL}
    |\Phi(t)\rangle &= e^{-i \phi(t)} |\tilde{\Phi}(t)\rangle
\end{align}
with
\begin{align}
\label{EQN:PHASE_SEP}
    |\tilde{\Phi}(t)\rangle &=  \hat{U}_\text{CS}e^{\hat{\kappa}(t)}  \ketz
\end{align}

The state $|\Phi(t)\rangle$ evolves according to the TD Schr\"{o}dinger equation and satisfies
\begin{align}
\label{EQN:TDSE}
     (\hat{H}(t) - i \partial_t )|\Phi(t)\rangle = 0
\end{align}
where $\partial_t = \frac{\partial}{\partial t}$, and the TD Hamiltonian is {given in Eq.~\ref{EQN:TD_HAMILTONIAN}.}
After the application of the time derivative, we find
\begin{align}
e^{-i \phi(t)} \left( \hat{H}(t) - \dot{\phi}(t) - i\partial_t \right) |\tilde{\Phi}(t)\rangle = 0
\end{align}
which implies that
\begin{align}
\dot{\phi}(t) |\tilde{\Phi}(t)\rangle =
( \hat{H}(t) - i\partial_t ) |\tilde{\Phi}(t)\rangle 
\end{align}
Left projection onto $\langle \tilde{\Phi}(t)|$
yields 
\begin{align}
     \dot{\phi}(t) = Q(t) = \langle \tilde{\Phi}(t)|( \hat{H}(t) - i\partial_t ) |\tilde{\Phi}(t)\rangle 
\end{align}
Here, we have introduced the TD quasi-energy, ${Q}(t)$,\cite{Hattig98_1} which is the time derivative of $\phi(t)$. {In the} absence of any external perturbation, $Q(t)$ reduces to the ground-state QED-HF energy. Expanding $|\tilde{\Phi}\rangle$, we have
\begin{align} 
\label{EQN:QUASI_ENERGY}
Q(t) &=  \braz e^{-\hat{\kappa}(t)} ( \hat{H}_\text{CS} + V(t)-i\partial_t) e^{\hat{\kappa}(t)}\ketz
\end{align}
Frequency-dependent response properties for our QED-HF state can now be determined from appropriate derivatives of the time-averaged quasi-energy,\cite{Hattig98_1}
\begin{align}
    \{Q\}_T = \frac{1}{T} \int_0^T dt~Q(t)
\end{align}

We can expand a TD observable, $\Omega_A (t)$, in terms of the Fourier components of the perturbing field as 
\begin{widetext}
\begin{align} 
    {\Omega_A(t) = \langle\Omega_A\rangle_0 + \sum_{\omega} \langle\langle \Omega_A; V(\omega)\rangle\rangle\, e^{-i\omega t} + \frac{1}{2}\sum_{\omega_1}\sum_{\omega_2} \langle\langle \Omega_A; V(\omega_1), V(\omega_2)\rangle\rangle\, e^{-i\omega_1 t} e^{-i\omega_2 t} + \cdots}
\end{align}
\end{widetext}
Here, $\langle \Omega_A \rangle_0$ represents the expectation value of the observable with respect to the time-independent state ({\em i.e.}, in the absence of the external perturbation). The double bracket notation is used to denote various response functions, which describe how the expectation value of $\hat{\Omega}_A$ responds to specific external perturbations. Following Ref.~\citenum{Hattig98_1}, the response functions can be identified as the following derivatives of the time-averaged quasi-energy:
\begin{align}
\label{EQN:ZEROTH_ORDER_RESPONSE_FUNCTION}
    \langle \Omega_A\rangle_0 &= \frac{\partial \{Q\}_T}{\partial \epsilon_A(0)} \\
    \label{EQN:FIRST_ORDER_RESPONSE_FUNCTION}
    \langle\langle \Omega_A; \Omega_B\rangle\rangle_{\omega_1} &= \frac{\partial^2 \{Q\}_T}{\partial \epsilon_A(\omega_0)\partial \epsilon_B(\omega_1)}; \\
    \omega_0 &= -\omega_1 \nonumber \\
\label{EQN:SECOND_ORDER_RESPONSE_FUNCTION}
    \langle\langle \Omega_A; \Omega_B, \Omega_C\rangle\rangle_{\omega_1,\omega_2} &= \frac{\partial^3 \{Q\}_T}{\partial \epsilon_A(\omega_0)\partial \epsilon_B(\omega_1)\partial\epsilon_C(\omega_2)}; \\
    \omega_0 &= -\omega_1 - \omega_2 \nonumber \\
    \text{etc.}\nonumber
\end{align}
Note that it is implied that the derivatives are evaluated at the limit that the perturbation strength is zero. 

In this work, we are concerned with the linear and quadratic response functions, which, by Wigner's $2n+1$ rule,\cite{Hylleraas30_209,Wigner35_477} can both be evaluated with knowledge of only the first-order response  of the wave function to the external perturbations.\cite{Jorgensen89_111} To derive the first-order response equations, we first expand the response parameters in orders of the perturbation
\begin{align}
    \hat{\kappa}(t) = \hat{\kappa}^{(0)} + \hat{\kappa}^{(1)}(t)  +\hat{\kappa}^{(2)}(t) +  ...
\end{align}
The zeroth-order parameters do not depend on the perturbation and vanish so that the quasi-energy recovers the ground-state energy in the absence of the perturbation. We can expand the first-order parameters in terms of the Fourier components of the perturbation and the perturbation strength as
\begin{align}
\label{EQN:RESPONSE_W}
    {\hat\kappa^{(1)}(t) = \sum_{\omega} e^{-i\omega t} \sum_A \epsilon_A(\omega)\, \hat\kappa_A^{(1)}(\omega)}
\end{align}
Like, $\hat{\kappa}(t)$, $\hat{\kappa}^{(1)}_A(\omega)$ can be expanded in terms of electron and photon degrees of freedom as
\begin{align}
\hat{\kappa}^{(1)}_A(\omega) =
& \sum_{ia} \left ( X^\text{e}_{A,ia}(\omega) \hat{a}^\dagger_a \hat{a}_i - Y^\text{e}_{A,ia}(\omega) \hat{a}^\dagger_i \hat{a}_a \right ) \nonumber \\
&+ X^\text{p}_A(\omega) \hat{b}^\dagger - Y^\text{p}_A(\omega) \hat{b}
\end{align}
where we have suppressed the perturbation order in the notation for the frequency-dependent response parameters on the right-hand side of this equation.

Like the response parameters, the time-averaged quasi-energy can be expanded in a perturbation series of the form
\begin{align}
\label{EQN:Q_PERT}
    \{Q\}_T = E_0 + \{Q\}_T^{(1)} + \{Q\}_T^{(2)} + ...
\end{align}
Following Ref.~\citenum{Hattig98_1}, the first-order response parameters are determined by making the second-order part of $\{Q\}_T$ stationary with respect to their variations. We isolate $\{Q\}_T^{(2)}$ by expanding $\{Q\}_T$ via the Baker--Campbell--Hausdorff series and collecting all terms of order two to obtain
\begin{widetext}
\begin{align}
\{Q\}_T^{(2)} = \frac{1}{T} \int_0^T dt \braz [V(t), \hat{\kappa}^{(1)}(t)] + \frac{1}{2} [[\hat{H}_{\text{CS}}, \hat{\kappa}^{(1)}(t)], \hat{\kappa}^{(1)}(t)] - \frac{i}{2} [\hat{\kappa}^{(1)}(t), \smash{\overset{\cdot}{\hat{\kappa}}}^{(1)}(t)] + [\hat{H}_{\text{CS}}, \hat{\kappa}^{(2)}(t)] - i\smash{\overset{\cdot}{\hat{\kappa}}}^{(2)}(t) \ketz
\end{align}
\end{widetext}
The single commutators $\braz [\hat{H}_{\text{CS}}, \hat{\kappa}^{(n)}(t)] \ketz$ with $n\in 1, 2$ both vanish. The electronic part vanishes because the QED-HF reference state is fully optimized and satisfies Brillouin's theorem. Similarly, it can easily be shown that the photon parts of these terms vanish because 
\begin{align}
\braz [\hat{H}_\text{CS},\hat{b}]\ketz = \braz [\hat{H}_\text{CS},\hat{b}^\dagger]\ketz  = 0
\end{align}
The term $\braz\smash{\overset{\cdot}{\hat{\kappa}}}^{(2)}(t)\ketz$ is also zero, and we are left with
\begin{widetext}
\begin{align}
\label{EQN:QT2}
\{Q\}_T^{(2)} = \frac{1}{T} \int_0^T dt \braz [V(t), \hat{\kappa}^{(1)}(t)] + \frac{1}{2} [[\hat{H}_{\text{CS}}, \hat{\kappa}^{(1)}(t)], \hat{\kappa}^{(1)}(t)] - \frac{i}{2} [\hat{\kappa}^{(1)}(t), \smash{\overset{\cdot}{\hat{\kappa}}}^{(1)}(t)] \ketz
\end{align}
\end{widetext}
By inserting the Fourier expansions of $V(t)$ (Eqs.~\ref{EQN:PERTURBATION_T} and \ref{EQN:PERTURBATION_W}) and $\hat{\kappa}^{(1)}(t)$ (Eq.~\ref{EQN:RESPONSE_W}) into Eq.~\ref{EQN:QT2} and evaluating the integral over time, one obtains
\begin{widetext}
\begin{align}
\label{EQN:Q2_W}
    \{Q\}_T^{(2)} = {\sum_\omega} \sum_{AB} \epsilon_A(\omega)\epsilon_B(-\omega) \braz [\hat{\Omega}_A, \hat{\kappa}_B^{(1)}(-\omega)] + \frac{1}{2} [[\hat{H}_{\text{CS}}, \hat{\kappa}_A^{(1)}(\omega)], \hat{\kappa}_B^{(1)}(-\omega)] + \frac{\omega}{2} [\hat{\kappa}_A^{(1)}(\omega), \hat{\kappa}_B^{(1)}(-\omega)]  \ketz
\end{align}
\end{widetext}
The response parameters at negative frequencies that appear in this expression [$\hat{\kappa}_B(-\omega)$] are related to the positive-frequency parameters via the anti-Hermiticity of $\hat{\kappa}(t)$. Specifically, we have
\begin{align}
X^\text{e}_{B,ia}(-\omega) &= [Y^\text{e}_{B,ia}(\omega)]^* = Y^\text{e}_{B,ia}(\omega)\\
Y^\text{e}_{B,ia}(-\omega) &= [X^\text{e}_{B,ia}(\omega)]^*  = X^\text{e}_{B,ia}(\omega)\\
X^\text{p}_B(-\omega) &= [Y^\text{p}_B(\omega)]^* = Y^\text{p}_B(\omega)\\
Y^\text{p}_B(-\omega) &= [X^\text{p}_B(\omega)]^* = X^\text{p}_B(\omega)
\end{align}
where we have assumed that $X^\text{e}_{A,ia}(\omega)=[X^\text{e}_{A,ia}(\omega)]^*$, etc.~because the reference state
{ and the perturbation operator considered in this work (\emph{i.e.}, the electric dipole operator) are real-valued. If the perturbation considered is complex-valued (\emph{e.g.}, magnetic-dipole operator), one has to explicitly consider the complex conjugate nature of excitation and de-excitation amplitudes.}

Equations to determine the first-order response amplitudes are obtained by making Eq.~\ref{EQN:Q2_W} stationary with respect to $X^\text{e}_{A,ia}(\omega)$, $Y^\text{e}_{A,ia}(\omega)$, $X^{\text{p}}_A(\omega)$, and $Y^\text{p}_A(\omega)$, which yields
\begin{widetext}
\begin{align}
\label{EQN:XE_RESPONSE}
\sum_{jb} (A_{ia, jb} - \delta_{ij}\delta_{ab}\omega )X^\text{e}_{C, jb}(\omega) + \sum_{jb}B_{ia, jb}Y^\text{e}_{C,jb}(\omega)  - \sqrt{\frac{ \omega_\text{cav}}{2}}\left ( {\bm \lambda} \cdot { \bm \mu}^\text{e}\right )_{ia}   \left ( X^\text{p}_C(\omega) + Y^\text{p}_C(\omega)\right )= {} & -\Omega_{C,ia}  \\
\label{EQN:YE_RESPONSE}
\sum_{jb} B_{ia, jb}X^\text{e}_{C,jb}(\omega) + \sum_{jb}(A_{ia, jb} + \delta_{ij}\delta_{ab}\omega )Y^\text{e}_{C,jb}(\omega)  - \sqrt{\frac{ \omega_\text{cav}}{2}}\left ( {\bm \lambda} \cdot { \bm \mu}^\text{e}\right )_{ia}   \left ( X^\text{p}_C(\omega) + Y^\text{p}_C(\omega)\right )= {} & -\Omega_{C, ia}  \\
\label{EQN:XP_RESPONSE}
- \sqrt{\frac{ \omega_\text{cav}}{2}} \sum_{ia} \left ( {\bm \lambda} \cdot { \bm \mu}^\text{e}\right )_{ia} \left ( X^\text{e}_{C,ia}(\omega) + Y^\text{e}_{C,ia}(\omega)\right ) + X^\text{p}_C(\omega) (\omega_\text{cav}-\omega) = {} & 0 \\
\label{EQN:YP_RESPONSE}
- \sqrt{\frac{ \omega_\text{cav}}{2}} \sum_{ia} \left ( {\bm \lambda} \cdot { \bm \mu}^\text{e}\right )_{ia} \left ( X^\text{e}_{C,ia}(\omega) + Y^\text{e}_{C,ia}(\omega)\right ) + Y^\text{p}_C(\omega) (\omega_\text{cav}+\omega) = {} & 0
\end{align}
\end{widetext}
In Eqs.~\ref{EQN:XE_RESPONSE} and \ref{EQN:YE_RESPONSE}, the symbols $A_{ia,jb}$ and $B_{ia, jb}$ represent elements of the {\bf A} and {\bf B} matrices of TD-HF theory, augmented by appropriate dipole self-energy contributions. For explicit expressions for the elements of these tensors, see Ref.~\citenum{DePrince22_9303}. { We have also introduced the matrix representation of the electric dipole operator, ${\bm \mu}^\text{e}$.} Note that the right-hand sides of Eqs.~\ref{EQN:XP_RESPONSE} and \ref{EQN:YP_RESPONSE} are zero because we have assumed that the perturbation operator involves only electronic degrees of freedom. 

To obtain an expression for the linear response function, we first rewrite Eqs.~\ref{EQN:XE_RESPONSE}--\ref{EQN:YP_RESPONSE} in matrix form as
\begin{align}
    {\bf M}(\omega){\bf Z}_C(\omega) = -{\bf V}_C
\end{align}
with 
\begin{align}
{\bf M}(\omega) =     \left ( \begin{array}{cccc}
    \bf{A} -\bf{1}\omega & \bf{B} & \bf{g} & \bf{g} \\
    \bf{B}  & \bf{A} +{\bf 1}\omega & \bf{g} & \bf{g} \\
    \bf{g}^\dagger & \bf{g}^\dagger &     \omega_\text{cav}-\omega & 0 \\
    \bf{g}^\dagger & \bf{{g}}^\dagger &    0  & \omega_\text{cav}+\omega
    \end{array} \right)
\end{align}
where ${\bf g}$ is a column vector with elements 
\begin{align}
g_{ia} = - \sqrt{\frac{ \omega_\text{cav}}{2}}\left ( {\bm \lambda} \cdot \hat{\bm \mu}^\text{e}\right )_{ia}
\end{align}
${\bf Z}_C(\omega)$ is a vector containing the response parameters
\begin{align}
\label{EQN:Z_C}
    {\bf Z}_C(\omega) = 
    \left( \begin{array}{c}
    \bm{X}^\text{e}_C(\omega) \\
    \bm{Y}^\text{e}_C(\omega) \\
    X^\text{p}_C(\omega) \\
    Y^\text{p}_C(\omega) 
    \end{array} \right)
\end{align}
and 
\begin{align}
\label{EQN:V_C}
    {\bf V}_C = 
    \left( \begin{array}{c}
    \bm{\Omega}_{C}^\text{ov} \\
    \bm{\Omega}_{C}^\text{ov} \\
    0 \\
    0 
    \end{array} \right)
\end{align}
Here, the superscript "ov" indicates that only the occupied-virtual block of the matrix representation of $\hat{\Omega}_C$ is used to fill the vector ${\bf V}_C$. We can now rewrite Eq.~\ref{EQN:Q2_W} in a much more compact form as
\begin{widetext}
\begin{align}
\label{EQN:Q2_W_COMPACT}
    \{Q\}_T^{(2)} = {} & {\sum_\omega} \sum_{AB} \epsilon_A(\omega)\epsilon_B(-\omega) [{\bf V}^\dagger_A {\bf Z}_B(\omega) + \frac{1}{2}{\bf Z}^\dagger_A(-\omega) {\bf M}(\omega){\bf Z}_B(\omega)] \nonumber \\
    {} = {} & {\sum_\omega} \sum_{AB} \epsilon_A(\omega)\epsilon_B(-\omega) [{\bf V}^\dagger_A {\bf Z}_B(\omega)  - \frac{1}{2}{\bf Z}^\dagger_A(-\omega) {\bf V}_B] \nonumber \\
    {} = {} & \frac{1}{2} {\sum_\omega} \sum_{AB} \epsilon_A(\omega)\epsilon_B(-\omega) {\bf V}^\dagger_A {\bf Z}_B(\omega)  
\end{align}
\end{widetext}
In the perturbation expansion of $\{Q\}_T$ in Eq.~\ref{EQN:Q_PERT}, the only term that is quadratic in the field strength is the second-order term. As such, 
\begin{align}
    \frac{\partial^2 \{Q\}_T}{\partial \epsilon_A(\omega_0)\partial \epsilon_B(\omega_1)} = \frac{\partial^2 \{Q\}_T^{(2)}}{\partial \epsilon_A(\omega_0)\partial \epsilon_B(\omega_1)}
\end{align}
and, from Eq.~\ref{EQN:Q2_W_COMPACT}, we can see that
\begin{align}
\label{EQN:LINEAR_RESPONSE}
    \langle\langle \Omega_A; \Omega_B\rangle\rangle_{\omega} = {\bf V}^\dagger_A {\bf Z}_B(\omega)
\end{align}
or 
\begin{align}
    \langle\langle \Omega_A; \Omega_B\rangle\rangle_{\omega} = \sum_{ia} \Omega_{A, ia} \left(X^\text{e}_{B,ia}(\omega) + Y^\text{e}_{B,ia}(\omega) \right)
\end{align}
Note that the linear response function does not explicitly depend on the first-order photon response because the external perturbation includes only electronic degrees of freedom. This is a general property of response functions involving only electronic degrees of freedom. In such cases, the photon response parameters only impact the response function indirectly, through the equations that define the electronic response parameters ({\em e.g.}, the first-order response equations, Eqs.~\ref{EQN:XE_RESPONSE} and \ref{EQN:YE_RESPONSE}). 

In this work, we are concerned with response functions involving the electric dipole as the observable and a perturbing electric field, so $V(\omega) = -\sum_A \epsilon_A(\omega) \hat{\mu}^{\text{e}}_A$. We can relate the electric dipole polarizability tensor to the linear response function
\begin{align}
\label{EQN:POL_LIN_RESPONSE}
    \alpha_{AB}(-\omega; \omega) = -\langle\langle \mu^\text{e}_A; \mu^\text{e}_B\rangle\rangle_{\omega}
\end{align}
and the first hyperpolarizability tensor to the quadratic response function
\begin{align}
    \beta_{ABC}(-\omega_1-\omega_2; \omega_1, \omega_2) = - \langle\langle \mu^\text{e}_A; \mu^\text{e}_{B}, \mu^\text{e}_{C}\rangle\rangle _{\omega_1, \omega_2}
\end{align}
Here, $A$, $B$, and $C$ represent Cartesian components of the electric field. By the $2n+1$ rule, the first hyperpolarizability tensor can be evaluated with only knowledge of the first-order response of the wave function, and, as noted above, because the perturbation operator for the hyperpolarizability contains only electronic degrees of freedom, we can use standard expressions for the quadratic response function presented elsewhere (see Ref.~\citenum{Hattig98_1}, for example).

From Eqs.~\ref{EQN:LINEAR_RESPONSE} and \ref{EQN:POL_LIN_RESPONSE}, the electric dipole polarizability takes the form
\begin{align}
\label{EQN:POLARIZABILITY_STANDARD}
    \alpha_{AB}(-\omega; \omega) = \sum_{ia} \mu^\text{e}_{A,ia} \left(X^\text{e}_{B,ia}(\omega) + Y^\text{e}_{B,ia}(\omega)\right)
\end{align}
where $X^\text{e}_{B,ia}(\omega)$ and $Y^\text{e}_{B,ia}(\omega)$ have been determined via Eqs.~\ref{EQN:XE_RESPONSE}--\ref{EQN:YP_RESPONSE} with $\Omega_{B,ia} = -\mu^\text{e}_{B,ia}$. 

Now, it is useful to note the following relationship
\begin{align}
\label{EQN:LALPHA}
{\sigma_B(\omega)} &= 
    \sum_A \lambda_A \alpha_{AB}(-\omega; \omega)\\
    &=\sum_{ia} ({\bm \lambda}\cdot \hat{\bm \mu}^\text{e})_{ia} \left(X^\text{e}_{B,ia}(\omega) + Y^\text{e}_{B,ia}(\omega)\right)
\end{align}
because the right-hand side appears in the equations for the first-order response of the photon part of the wave function (assuming the response parameters correspond to an electric dipole perturbation). Inserting the left-hand side of Eq.~\ref{EQN:LALPHA} into Eqs.~\ref{EQN:XP_RESPONSE} and \ref{EQN:YP_RESPONSE} and rearranging leads to compact expressions for the photon response parameters:
\begin{align}
    X^\text{p}_C(\omega) &= \sqrt{\frac{ \omega_\text{cav}}{2}} \frac{{\sigma_C(\omega)}}{\omega_\text{cav}-\omega} \\
    Y^\text{p}_C(\omega) &= \sqrt{\frac{ \omega_\text{cav}}{2}} \frac{{\sigma_C(\omega)}}{\omega_\text{cav}+\omega}
\end{align}
We can now insert these definitions into the equations for the first-order response of the electron part of the wave function to an electric dipole perturbation to obtain
\begin{widetext}
\begin{align}
\label{EQN:XE_RESPONSE_COMPACT}
\sum_{jb} (A_{ia, jb} - \delta_{ij}\delta_{ab}\omega )X^\text{e}_{C,jb}(\omega) + \sum_{jb}B_{ia, jb}Y^\text{e}_{C,jb}(\omega) = \mu^\text{e}_{C,ia} +  {\tau(\omega)\sigma_C(\omega)}\left ( {\bm \lambda} \cdot \hat{\bm \mu}^\text{e}\right )_{ia}  \\ 
\label{EQN:YE_RESPONSE_COMPACT}
\sum_{jb} B_{ia, jb}X^\text{e}_{C,jb}(\omega) + \sum_{jb}(A_{ia, jb} + \delta_{ij}\delta_{ab}\omega )Y^\text{e}_{C,jb}(\omega)  = \mu^\text{e}_{C,ia} +  {\tau(\omega)\sigma_C(\omega)}\left ( {\bm \lambda} \cdot \hat{\bm \mu}^\text{e}\right )_{ia}  
\end{align}
\end{widetext}
{where we have also introduced the symbol
\begin{align}
    \tau(\omega) = \frac{\omega_\text{cav}^2}{\omega_\text{cav}^2 - \omega^2}
\end{align}
}In this form, the QED-HF first-order response equations more closely resemble the standard HF ones, except that the dipole matrix elements on the right-hand side have been augmented by the term involving the ${\bm \lambda}$-weighted polarizability. This form also provides insight into when the cavity mode will induce significant changes in the electron response, {\em i.e.}, when there is feedback between the electric dipole polarizability and the electronic response parameters. 

{
Now that we have identified the compact form of the electronic response equations in Eqs.~\ref{EQN:XE_RESPONSE_COMPACT} and \ref{EQN:YE_RESPONSE_COMPACT}, we are able to show that cavity-induced changes to the electric-dipole polarizability are separable into contributions arising from the DSE and BLC terms. First, let us cast Eqs.~\ref{EQN:XE_RESPONSE_COMPACT} and \ref{EQN:YE_RESPONSE_COMPACT} in matrix form:
\begin{align}
\label{EQN:response_most_compact}
    {\bf E}(\omega) {\bf Z}_C(\omega) = {\bf d}_C + \tau(\omega)\sigma_C(\omega) {\bf W}
\end{align}
where
\begin{align}
{\bf E}(\omega) =     \left ( \begin{array}{cc}
    \bf{A} -\bf{1}\omega & \bf{B} \\
    \bf{B}  & \bf{A} +{\bf 1}\omega
    \end{array} \right)
\end{align}
\begin{align}
    {\bf W} = 
    \left( \begin{array}{c}
    ({\bm \lambda}\cdot\hat{\bm{\mu}}^\text{e})^\text{ov} \\
    ({\bm \lambda}\cdot\hat{\bm{\mu}}^\text{e})^\text{ov} 
    \end{array} \right)
\end{align}
and
\begin{align}
    {\bf d}_C = 
    \left( \begin{array}{c}
    \bm{\mu}_{C}^\text{e,ov} \\
    \bm{\mu}_{C}^\text{e,ov} 
    \end{array} \right)
\end{align}
Here, we have truncated the response vector, ${\bf Z}_C$, to include only the electronic part, {\em i.e.}
\begin{align}
    {\bf Z}_C(\omega) = 
    \left( \begin{array}{c}
    \bm{X}^\text{e}_C(\omega) \\
    \bm{Y}^\text{e}_C(\omega)
    \end{array} \right)
\end{align}
The electronic response can be determined from Eq.~\ref{EQN:response_most_compact} as
\begin{align}
    {\bf Z}_C(\omega) &= {\bf E}(\omega)^{-1} {\bf d}_C + \tau(\omega)\sigma_C(\omega) {\bf E}(\omega)^{-1}{\bf W} \nonumber \\
    &= {\bf Z}_C^0(\omega) + \tau(\omega)\sigma_C(\omega)Z_\lambda^0(\omega)
\end{align}
Now, we can express the polarizability as
\begin{align}
    \alpha_{DC}(-\omega;\omega) &= {\bf d}_D^\dagger {\bf E}(\omega)^{-1} {\bf d}_C \nonumber \\
    &+ \tau(\omega)\sigma_C(\omega) {\bf d}_D^\dagger {\bf E}(\omega)^{-1}{\bf W} \nonumber \\
    &= \alpha_{DC}^0(\omega) + \tau(\omega)\sigma_C(\omega)\alpha_{D,\lambda}^0(\omega)
\end{align}
Here, $\alpha_{DC}^0(\omega)$ is a polarizability-like quantity associated with the electronic response determined in the absence of any BLC term, {\em e.g.}, by solving ${\bf E}(\omega){\bf Z}^0_C(\omega) = {\bf d}_C$, but still accounting for the DSE contributions to the {\bf A} and {\bf B} matrices that enter ${\bf E}(\omega)$. Similarly, $\alpha_{D,\lambda}^0(\omega)$, resembles a linear-response function associated with the perturbation, ${\bf W}$.
However, interestingly, $\alpha_{D,\lambda}^0(\omega)$ can be determined directly from $\alpha_{DC}^0(\omega)$, without solving a separate response problem. If we note that
\begin{align}
    {\bm W} = \sum_E \lambda_E {\bm d}_E
\end{align}
then 
\begin{align}
    \alpha_{D,\lambda}^0(\omega) &= {\bf d}_D^\dagger {\bf E}(\omega)^{-1}\sum_E \lambda_E {\bm d}_E \nonumber \\
    &=  \sum_E \lambda_E \alpha_{DE}^0(\omega)
\end{align}
We introduce a BLC-free analogue of the ${\bm \lambda}$-weighted polarizability 
\begin{align}
    \sigma_C^0(\omega) = \sum_D \lambda_D \alpha_{DC}^0(\omega)
\end{align}
as well as the quantity
\begin{align}
    \gamma(\omega) = \sum_D  \lambda_D \alpha_{D,\lambda}^0(\omega) = \sum_{DE} \lambda_D \lambda_E \alpha^0_{DE}(\omega)
\end{align}
after which we can express the ${\bm \lambda}$-weighted polarizability as
\begin{align}
    \sigma_C(\omega) 
    &=\sigma_C^0(\omega) + \tau(\omega)\sigma_C(\omega)\sum_D \lambda_D \alpha_{D,\lambda}^0(\omega) \nonumber \\
    & = \sigma_C^0(\omega) + \tau(\omega)\gamma(\omega)\sigma_C(\omega)
\end{align}
Isolating $\sigma_C(\omega)$, we have
\begin{align}
    \sigma_C(\omega) = \frac{\sigma_C^0(\omega)}{1-\tau(\omega)\gamma(\omega)}
\end{align}
which we can substitute back into the expression for the polarizability itself to obtain
\begin{align}
\label{EQN:POLARIZABILITY_CROSSING}
    \alpha_{DC}(-\omega;\omega) = \alpha_{DC}^0(\omega) + \frac{\tau(\omega)\sigma_C^0(\omega)\alpha_{D,\lambda}^0(\omega)}{1-\tau(\omega)\gamma(\omega)}
\end{align}
Here, we have the remarkable result that the total polarizability can be determined from response vectors calculated in the absence of the BLC term, {\em i.e.}, from ${\bf Z}_C^0(\omega)$. This result will be useful for interpreting cavity-induced changes to the polarizability, which can be decomposed in the same way. Before moving on, we emphasize that, in Sec.~\ref{sec:results_discussion}, the polarizability values we report are evaluated according to Eq.~\ref{EQN:POLARIZABILITY_STANDARD}, with the electronic response determined according to Eqs.~\ref{EQN:XE_RESPONSE_COMPACT} and \ref{EQN:YE_RESPONSE_COMPACT}. The form of the polarizability in Eq.~\ref{EQN:POLARIZABILITY_CROSSING} is equivalent (we have verified this equivalence numerically) and will be useful in interpreting the results of our calculations.
}

\subsection{RT-TD-QED-HF Theory}

Let us now consider an alternative approach to obtain dynamic QED-HF response properties through RT-TD simulations. We begin by recalling that the TD quasi-energy for the phase-separated wave function defined by Eqs.~\ref{EQN:TD_WFN_FINAL} and \ref{EQN:PHASE_SEP} was derived from the TD Schr\"{o}dinger equation (Eq.~\ref{EQN:TDSE}) with the TD Hamiltonian defined by Eq.~\ref{EQN:TD_HAMILTONIAN}. Let us instead defined the phase-separated wave function as
\begin{align}
\label{EQN:TD_WFN_FINAL_AGAIN}
    |\Phi(t)\rangle &= e^{-i \phi(t)} |\tilde{\Phi}(t)\rangle
\end{align}
with
\begin{align}
\label{EQN:PHASE_SEP_AGAIN}
    |\tilde{\Phi}(t)\rangle &=  e^{\hat{\kappa}(t)}  \ketz
\end{align}
({\em i.e.}, without the coherent-state transformation operator). We can evolve this state according to a modified time-dependent Schr\"{o}dinger equation involving the CS-transformed Hamiltonian so that
\begin{align}
\label{EQN:TDSE2}
     (\hat{H}_\text{CS} + V(t) - i \partial_t )|\Phi(t)\rangle = 0
\end{align}
Now, following the same procedure we used to derive the time-dependent quasi-energy in Sec.~\ref{SEC:RESPONSE_THEORY}, we obtain exactly the same expression for the time-dependent quasi-energy in Eq.~\ref{EQN:QUASI_ENERGY}. This result suggests that the RT-TD problem can be simplified by working with a TD wave function of the form defined in Eq.~\ref{EQN:PHASE_SEP_AGAIN}, with dynamics governed by the CS-transformed Hamiltonian. 

We must now develop the equations of motion (EOMs) for the QED-HF electron and photon density matrices. We define the one-electron reduced density matrix (1RDM) for the time-dependent state as
\begin{align}
    D_{pq}^\text{e}(t) = \langle \Phi(t)| \hat{a}^\dagger_q \hat{a}_p |\Phi(t)\rangle
\end{align}
with $\Phi(t)$ defined by Eqs.~\ref{EQN:TD_WFN_FINAL_AGAIN} and \ref{EQN:PHASE_SEP_AGAIN}. Here, the indices $p$ and $q$ refer to general molecular spin orbitals (occupied or unoccupied). We define the photon density matrix as 
\begin{align}
    D^\text{p}_{XY}(t) = \langle \Phi(t)| Y^\text{p}\rangle\langle X^\text{p} |\Phi(t)\rangle
\end{align}
where $| Y^\text{p}\rangle\langle X^\text{p} |$ is the photon density operator. Note that the photon space is restricted to span only $|0^\text{p}\rangle$ and $|1^\text{p}\rangle$. With this restriction, we can write
\begin{align}
    \hat{b} &= |0^\text{p}\rangle \langle 1^\text{p}| \\
    \hat{b}^\dagger &= |1^\text{p}\rangle \langle 0^\text{p}|
\end{align}
and
\begin{align}
    \hat{b}^\dagger \hat{b} &= |1^\text{p}\rangle \langle 1^\text{p}|
\end{align}

The { EOMs} for ${\bf D}^\text{e}$ and ${\bf D}^\text{p}$ can be { derived from the TD variational principle, which yields the} commutator of the appropriate density operator and the time-dependent CS-transformed Hamiltonian,
\begin{align}
    \hat{H}_\text{CS}(t) = \hat{H}_\text{CS} - {\bm \epsilon}(t)\cdot \hat{{\bm \mu}}^\text{e}
\end{align}
where we have chosen $V(t) = - {\bm \epsilon}(t)\cdot \hat{{\bm \mu}}^\text{e}$, and ${\bm \epsilon}(t)$ represents a time-dependent electric field.
The EOM for the 1RDM is
\begin{align}
    \dot{D}_{pq}^\text{e}(t) = -i \langle \Phi(t) | [\hat{a}^\dagger_q \hat{a}_p, \hat{H}_\text{CS}(t)]|\Phi(t)\rangle
\end{align}
which can be expanded as
\begin{align}
    \dot{D}_{pq}^\text{e}(t) = &-i \langle \Phi(t) | [\hat{a}^\dagger_q \hat{a}_p, \hat{H}_\text{e} + \hat{H}_\text{DSE}]|\Phi(t)\rangle \nonumber \\
    &+i \langle \Phi(t) | [\hat{a}^\dagger_q \hat{a}_p, {\bm \epsilon}(t)\cdot \hat{{\bm \mu}}^\text{e}]|\Phi(t)\rangle \nonumber \\
    &+i \sqrt{\frac{\omega_\text{cav}}{2}}\langle \Phi(t) | [\hat{a}^\dagger_q \hat{a}_p, {\bm \lambda}\cdot \hat{\bm{\mu}}^\text{e}  ] (\hat{b}^\dagger + \hat{b})|\Phi(t)\rangle
\end{align}
where we have introduce shorthand notation for the DSE part of the Hamiltonian
\begin{align}
    \hat{H}_\text{DSE} = \frac{1}{2} \left({\bm \lambda}\cdot [\hat{\bm{\mu}}^\text{e} - \langle \hat{\bm{\mu}}^\text{e}\rangle ]\right)^2
\end{align}
After some tedious manipulations { (see the Supporting Information for more details)}, we arrive at
\begin{align}
\label{EQN:DE_EOM}
    \dot{D}_{pq}^\text{e}(t) = &-i [{\bf F}(t) - {\bm \epsilon}(t)\cdot{ \bm \mu}^\text{e}, {\bf D}^\text{e}(t)]_{pq} \nonumber \\
    &+i \sqrt{\frac{\omega_\text{cav}}{2}} (D_{01}^\text{p}{ (t)} + D_{10}^\text{p}{ (t)}) [{\bm \lambda}\cdot{ \bm \mu}^\text{e}, {\bf D}^\text{e}(t)]_{pq} 
\end{align}
Here, we have introduced the TD Fock matrix, $\mathbf{F}(t)$, the elements of which are given by
\begin{align}
\label{eqn:fock_0}
    F_{pq}(t) = {} & h_{pq} - \left( {\bm{\lambda}}\cdot \langle {\bm{\hat{\mu}}}^\text{e}\rangle \right) ({\bm \lambda}\cdot { \bm \mu}^\text{e})_{pq} - \frac{1}{2} q_{pq} \nonumber \\
    & + J_{pq}(t) - K_{pq}(t) + J_{pq}^\text{DSE}(t) - K_{pq}^\text{DSE}(t)
\end{align}
The symbol $h_{pq}$ refers to an element of the core Hamiltonian matrix, and $J_{pq}(t)$ and $K_{pq}(t)$ are the usual Coulomb and exchange matrices, respectively, defined as
\begin{align}
    J_{pq}(t) = \sum_{rs} (pq|rs) D^\text{e}_{sr}(t) \\
    K_{pq}(t) = \sum_{rs}  (ps|rq) D^\text{e}_{sr}(t)
\end{align}
where the symbol $(pq|rs)$ represents an electron repulsion integral in chemists' notation. 
The symbols $J_{pq}^\text{DSE}(t)$ and $K_{pq}^\text{DSE}(t)$ represent Coulomb- and exchange-like terms that derive from the DSE part of the Hamiltonian and are defined as
\begin{align}
\label{EQN:JDSE}
    J^\text{DSE}_{pq}(t) & = ({\bm \lambda}\cdot { \bm \mu}^e)_{pq} ~{\bm \lambda}\cdot \langle \hat{\bm \mu}^\text{e}(t) \rangle  \\
\label{EQN:KDSE}
    K^\text{DSE}_{pq}(t) & = \sum_{rs} ({\bm \lambda}\cdot { \bm \mu}^e)_{ps} ({\bm \lambda}\cdot { \bm \mu}^e)_{rq} D^\text{e}_{sr}(t) 
\end{align}
In Eq.~\ref{EQN:JDSE}, we have introduced the TD electric dipole moment,  $\langle \hat{\bm \mu}^\text{e}(t)\rangle$, which is evaluated as the trace of the matrix representation of the electric dipole operator against ${\bf D}^\text{e}(t)$:
\begin{equation}
\label{eqn:td_dipole}
    \expval{\hat{\bm \mu}^\text{e}(t)} = \Tr[{ \bm \mu}^\text{e}\mathbf{D}^\text{e}(t)].
\end{equation}
In Eq.~\ref{eqn:fock_0}, the symbol $q_{pq}$ represents a dressed electric quadrupole integral that also derives from $\hat{H}_\text{DSE}$ and is defined by
\begin{align}
        q_{pq} &= - \sum_{AB \in \{x,y,z\}} \lambda_A \lambda_B \int \phi^*_p r_A r_B \phi_{q} d\tau
\end{align}
Here, $\phi_q$ is a molecular spin orbital, and $r_A$ is a Cartesian component of the position vector. 

The EOM for the photon density matrix is
\begin{align}
    \dot{D}_{XY}^\text{p}(t) = -i \langle \Phi(t) | [|Y^\text{p}\rangle\langle X^\text{p}|, \hat{H}_\text{CS}(t)]|\Phi(t)\rangle
\end{align}
Expanding $\hat{H}_\text{CS}(t)$ leads to
\begin{widetext}
\begin{align}
    \dot{D}_{XY}^\text{p}(t) = -i \omega_\text{cav}\langle \Phi(t) | [|Y^\text{p}\rangle\langle X^\text{p}|, \hat{b}^\dagger\hat{b}]|\Phi(t)\rangle
    + i \sqrt{\frac{\omega_\text{cav}}{2}}\langle \Phi(t) | {\bm \lambda}\cdot (\hat{\bm{\mu}}^\text{e} - \langle \hat{\bm{\mu}}^\text{e} \rangle)[|Y^\text{p}\rangle\langle X^\text{p}|,  (\hat{b}^\dagger + \hat{b})]|\Phi(t)\rangle
\end{align}
\end{widetext}
or, in matrix notation,
\begin{align}
\label{EQN:DP_EOM}
    \dot{D}_{XY}^\text{p}(t) = &-i [{\bf H}^\text{p} + {\bf H}^\text{ep}(t), {\bf D}^\text{p}(t)]_{XY} 
\end{align}
In Eq.~\ref{EQN:DP_EOM}, we have introduced the matrix representations of the bare photon Hamiltonian, $\omega_\text{cav}\hat{b}^\dagger \hat{b}$,
\begin{align}
    {\bf H}^\text{p} = \begin{pmatrix}
        0 & 0 \\
        0 & \omega_\text{cav} \\
    \end{pmatrix}
\end{align}
and the {BLC} term
\begin{align}
    {\bf H}^\text{ep}(t) = -\sqrt{\frac{\omega_\text{cav}}{2}} {\bm \lambda}\cdot (\langle \hat{\bm \mu}^\text{e}(t)\rangle - \langle \hat{\bm{\mu}}^\text{e} \rangle)
    \begin{pmatrix}
        0 & 1 \\
        1 & 0 \\
    \end{pmatrix}
\end{align}
{ Before moving on, it is worth mentioning here that an alternative derivation of the HF response theory outlined in the previous subsection could have started from the EOMs for QED-TD-HF derived in this section, rather than relying on time-averaged quasi-energy (see, \emph{e.g.}, Refs.~\citenum{Jorgensen85_3235,Jorgensen95_857}).}

\subsection{Dynamic Response Properties from RT-TD-QED-HF Propagations}

The procedure to obtain response properties from RT simulations has been outlined in Refs.~\citenum{Li13_064104,Pedersen23_154102,Crawford25_1908}. Following those works, we apply a cosine-shaped field with frequency $\omega$ and a quadratic ramping envelope over the first $n$ periods,\cite{Pedersen23_154102,Crawford25_1908}
\begin{equation}
\label{eqn:td_field}
    \boldsymbol{\epsilon}(t) =
    \begin{cases}
        \frac{2t^2}{t_\text{r}^2} \boldsymbol{\epsilon} \cos(\omega t) & t < \frac{t_\text{r}}{2} \\
        \left[ 1 - \frac{2(t-t_\text{r})^2}{t_\text{r}^2} \right] \boldsymbol{\epsilon} \cos(\omega t) & \frac{t_\text{r}}{2} \leq t < t_\text{r} \\
        \boldsymbol{\epsilon} \cos(\omega t) & t \geq t_\text{r} \\
    \end{cases}
\end{equation}
in which $t_\text{r} = n2\pi/\omega$ is the ramping duration and $\boldsymbol{\epsilon}$ is the vector of field amplitudes. Quadratic ramping produces a smooth first derivative of the envelope at $t=0$ and $t=t_\text{r}$ compared to the linear ramping employed in Ref.~\citenum{Li13_064104}, which could induce high-frequency oscillations in the dipole moment.\cite{Pedersen23_154102,Crawford25_1908}

As shown in Ref.~\citenum{Li13_064104}, the TD dipole moment can be expressed as the expansion
\begin{align}
\label{eqn:td_dipole_taylor}
    \mu_{A}^\text{e}(t) = {} & \expval*{\mu_A^\text{e}}_{0} 
               + \sum_{B} \mu_{A B}^{\text{e},(1)}(t) \epsilon_{B}
               + \frac{1}{2} \sum_{B C} \mu_{A B C}^{\text{e},(2)}(t) \epsilon_{B} \epsilon_{C} \nonumber \\
               & {} + \frac{1}{6} \sum_{B C D} \mu_{A B C D}^{\text{e},(3)}(t) \epsilon_{B} \epsilon_{C} \epsilon_{D}
               + \ldots
\end{align}
where the subcripts $A, B, C, D$ are Cartesian coordinate indices, $\expval*{\mu_A^\text{e}}_{0}$ is a Cartesian component of the field-free dipole moment, and $\epsilon_B$, etc.~are Cartesian components of the field strength, ${\bm \epsilon}$ in Eq.~\ref{eqn:td_field}. A variety of dynamic response properties can be derived from the coefficient expansions using the following relationships\cite{Cotter90_12}
\begin{equation}
\label{eqn:mu_1}
    \mu_{A B}^{\text{e},(1)}(t) = \alpha_{A B}(-\omega;\omega) \cos(\omega t)
\end{equation}
\begin{equation}
\label{eqn:mu_2}
    \mu_{A B C}^{\text{e},(2)}(t) = \frac{1}{4}[\beta_{A B C}(-2\omega;\omega,\omega) \cos(2\omega t) + \beta_{A B C}(0;\omega,-\omega)]
\end{equation}
\begin{align}
\label{eqn:mu_3}
    \mu_{A B C D}^{\text{e},(3)}(t) = {} & \frac{1}{24}[\gamma_{A B C D}(-3\omega;\omega,\omega,\omega) \cos(3\omega t) \nonumber \\
                        & {} + 3\bar{\gamma}_{A B C D}(-\omega;\omega,\omega,-\omega) \cos(\omega t)]
\end{align}
and so on. In Eqs~\ref{eqn:mu_1}--\ref{eqn:mu_3}, $\alpha_{A B}(-\omega;\omega)$ is the frequency-dependent electric dipole polarizability, $\beta_{A B C}(-2\omega;\omega,\omega)$ and $\beta_{A B C}(0;\omega,-\omega)$ are the frequency-dependent first hyperpolarizabilities related to second harmonic generation (SHG) and optical rectification (OR), respectively, and $\gamma_{A B C D}(-3\omega;\omega,\omega,\omega)$ and $\bar{\gamma}_{A B C D}(-\omega;\omega,\omega,-\omega)$ are the frequency-dependent second hyperpolarizabilities related to third harmonic generation and degenerate four-wave mixing, respectively. As mentioned earlier, this work focuses on the dynamic polarizability (Eq.~\ref{eqn:mu_1}) and first hyperpolarizabilities (Eq.~\ref{eqn:mu_2}).

In order to extract the frequency-dependent response properties from Eq.~\ref{eqn:td_dipole_taylor}, we apply finite-field differentiation to this equation by computing the TD dipole moment at different field strengths and polarizations. For a $B$-polarized field with strength $\pm m \epsilon_{B}$ ($m=1,2,3,\ldots$), Eq.~\ref{eqn:td_dipole_taylor} becomes
\begin{align}
\label{eqn:td_dipole_field_str}
    \mu_{A}^\text{e}(t, \pm m \epsilon_{B}) = {} & \expval*{\mu_A^\text{e}}_{0}
                            \pm \mu_{A B}^{\text{e},(1)}(t) m \epsilon_{B}
                            + \mu_{A B B}^{\text{e},(2)}(t) m^{2} \epsilon_{B}^{2} \nonumber \\
                            & {} \pm \mu_{A B B B}^{\text{e},(3)}(t) m^{3} \epsilon_{B}^{3}
                            + \ldots
\end{align}
where we have added the field dependence to the notation for the TD dipole moment on the left-hand side of this equation.
We can then form the symmetric and antisymmetric linear combinations
\begin{equation}
\label{eqn:delta_m_pm}
    \Delta\mu_{A B}^{\text{e},m\pm}(t) = \mu_{A}^\text{e}(t, m \epsilon_{B}) \pm \mu_{A}^\text{e}(t, -m \epsilon_{B})
\end{equation}
which subsequently are used to determine the different terms in the Taylor expansion by eliminating higher-order contributions. For example, using $m=1$ and $m=2$, we obtain
\begin{equation}
\label{eqn:mu_1_3rd_order}
    \mu_{A B}^{\text{e},(1)}(t) = \frac{8 \Delta\mu_{A B}^{\text{e},1-}(t) - \Delta\mu_{A B}^{\text{e},2-}(t)}{12 \epsilon_{B}} + \mathscr{O}(\epsilon_{B}^{4})
\end{equation}
and
\begin{equation}
\label{eqn:mu_2_3rd_order}
    \mu_{A B B}^{\text{e},(2)}(t) = \frac{16 \Delta\mu_{A B}^{\text{e},1+}(t) - \Delta\mu_{A B}^{\text{e},2+}(t) - 30 \expval*{\mu_A^\text{e}}_{0}}{24 \epsilon_{B}^{2}} + \mathscr{O}(\epsilon_{B}^{4})
\end{equation}
that are correct through third order in the Taylor expansion. By fitting Eq.~\ref{eqn:mu_1_3rd_order} to Eq.~\ref{eqn:mu_1}, one obtains $\alpha_{A B}(-\omega;\omega)$ as the amplitude of a cosine function with the frequency $\omega$. Similarly, by fitting $\mu_{A B B}^{\text{e},(2)}(t)$ in Eq.~\ref{eqn:mu_2_3rd_order} to Eq.~\ref{eqn:mu_2}, one obtains a shifted oscillation with frequency $2\omega$, in which the amplitude and shift are $\frac{1}{4} \beta_{A B B}(-2\omega;\omega,\omega)$ and $\frac{1}{4} \beta_{A B B}(0;\omega,-\omega)$, respectively.

\section{Computational Details}
\label{sec:comp_details}

We validate our QED-HF response-theory implementation against response properties extracted from TD-QED-HF simulations in Sec.~\ref{sec:results:tdhf}. For this numerical verification study, components of the dynamic electric dipole polarizability and hyperpolarizability tensors were evaluated for the formaldehyde molecule with the aug-cc-pVDZ basis set. \cite{Dunning89_1007} The molecular geometry was optimized at the restricted HF level within the same basis while enforcing $C_{2v}$ symmetry. For RT-TD- and response-theory-based QED-HF simulations, the molecule was oriented in the $yz$-plane, with the dipole moment aligned along the $z$-direction. For TD-QED-HF simulations, the time integration was carried out using a standard 4th-order Runge--Kutta (RK4) integrator (with a time step of $\dd t = 0.1$ a.u.~[24 as]), the external field strengths were chosen to be { $\pm0.001$} and {$\pm0.002$} a.u.~in each Cartesian direction, and the field was ramped quadratically according to Eq.~\ref{eqn:td_field} over 10 periods. We considered two perturbing frequencies: 0.0428 a.u.~($\approx 1065$ nm) and 0.0656 a.u.~($\approx 695$ nm). The fittings involving Eqs.~\ref{eqn:mu_1_3rd_order} and \ref{eqn:mu_2_3rd_order} were carried out over 4 periods of the applied field's frequency.
{ The field strength and numbers of periods for ramping and fitting were chosen to obtain numerically converged response tensors (see the Supporting Information for additional data testing these parameters).}
We used cavity coupling constants ranging from 0.00 to 0.05 a.u.~at 0.01 a.u.~intervals. Note that $\lambda=0.05$ a.u.~corresponds to a cavity volume of about 0.74 nm$^3$, which falls under the ``picocavity'' category.\cite{Baumberg22_5859} In all RT-TD- and response-theory-based QED-HF calculations in this validation study, the cavity was polarized along the $z$ axis (aligned with the dipole moment of the formaldehyde molecule), and the cavity frequency was chosen to be resonant with a bright $A_1$-symmetry electronically excited state ($\omega_\text{cav} = 0.354723$ a.u.).

After numerically verifying the QED-HF response theory implementation against TD-QED-HF simulations, we examine cavity-induced changes in response properties in several molecular systems, namely, formaldehyde, urea, and pNA described by the d-aug-cc-pVTZ basis set.\cite{Dunning89_1007,Harrison92_6796,Dunning94_2975} As shown in Ref.~\citenum{DePrince25_8024}, this basis set is sufficiently { large} to provide numerically converged static polarizabilities and first hyperpolarizabilities at the QED-HF level of theory. Structures of formaldehyde, urea, and pNA were optimized at the restricted HF level of theory, using the same basis set. For formaldehyde and pNA, we employed $C_{2v}$ symmetry with the molecules lying on the $yz$ plane, whereas, for urea, we enforce $C_{2}$ symmetry with the principal axis aligned along the $z$-direction. In each case, the molecular dipole moment was aligned along the $z$-direction. Response properties were obtained for perturbing frequencies in the range $\omega=0.04$--0.20 a.u.~(near infra-red to deep ultraviolet wavelength range) and coupling constants spanning $\lambda=0.00$--0.05 a.u., with 0.01 a.u.~increments in both quantities. The cavity frequency is chosen such that it is matches the excitation energy of a bright $A_1$-symmetry (for formaldehyde and pNA) or $A$-symmetry (for urea) state. Table \ref{tab:vee_and_os} provides the relevant excitation energies and associated oscillator strengths. Note that we use the same cavity frequency for calculations where the cavity mode is polarized along the $x$-, $y$- and $z$-directions.

{ We also computed (QED-)HF dynamic polarizability tensors of the formate anion using the d-aug-cc-pVTZ basis set. We performed two sets of calculations where the center of mass of the anion is placed at the origin and at (10 \AA, 0 \AA, 0 \AA) in order to demonstrate the origin-invariance of the response tensors. The results of these calculations are provided in the Supporting Information.}

\begin{table}[!htbp]
    \centering
    \caption{Cavity-free vertical excitation energies (VEE, in $E_\text{h}$ and eV) and oscillator strengths of bright $A_1$-symmetry (for formaldehyde and pNA) or $A$-symmetry (for urea) states. }
    \label{tab:vee_and_os}
    \begin{tabular*}{\linewidth}{@{\extracolsep{\fill}} cccc}
        \hline\hline
        \multirow{2}{*}{Molecule} & \multicolumn{2}{c}{VEE} & \multirow{2}{*}{Oscillator strength} \\
        \cline{2-3}
        & $E_\text{h} $ & eV & \\
        \hline
        Formaldehyde & 0.356361 & 9.697 & 0.1152 \\
        Urea         & 0.331727 & 9.027 & 0.0876 \\
        pNA          & 0.258330 & 7.030 & 0.4990 \\
        \hline\hline
    \end{tabular*}
\end{table}

All response-theory-based calculations were performed with the Hilbert\cite{hilbert} package, which is a plugin to \textsc{Psi4},\cite{Sherrill20_184108} with an unrestricted HF reference. The RT-TD-\mbox{(QED-)HF} simulations were performed {using the DUMB Polaritonic HF Python} code,{\cite{dumbcode}} with atomic integrals taken from \textsc{Psi4}. These implementations were verified numerically at the zero coupling limit against the dynamic response properties computed using GAMESS.\cite{Gordon20_154102} Calculations within the aug-cc-pVDZ basis set made use of 4-index electron repulsion integrals (ERIs), whereas calculations carried out within the d-aug-cc-pVTZ basis set employed the density fitting { (DF)} approximation to the ERIs and the def2-universal-JKFIT auxiliary basis set,\cite{Weigend08_167} which is the default JK-type auxiliary basis set for d-aug-cc-pVTZ in \textsc{Psi4}.
{ Additional tests for the formaldehyde molecule showed that the DF approximation using def2-universal-JKFIT basis set produces dynamic polarizabilities, OR, and SHG tensors with quantitatively similar patterns to those obtained with explicit 4-index ERIs (using the same geometry and cavity frequency and $z$-polarized cavity). For example, the error resulting from the DF approximation with the def2-universal-JKFIT basis set is on the order of 0.001--0.003 a.u.~for dynamic polarizabilities and 0.011--0.158 a.u.~for the OR and SHG tensors, which translate to a maximum discrepancy of 0.7\%.}

{ In all of our response-theory-based calculations, the energy convergence threshold was set at $10^{-10}$ a.u. The density and response equation convergence threshold was set at $10^{-10}$ a.u.~as well for formaldehyde, urea, and formate anion, and at $10^{-8}$ a.u.~for pNA. In a few cases where the response equations were difficult to converge (e.g., at $\omega$ values approaching a pole in the response function), we relaxed the density and response equation convergence threshold to as high as $6\times10^{-7}$ a.u.; we note that the resulting (QED-)HF energies are still within $10^{-10}$ a.u.~from the calculations with the tighter threshold. The $10^{-10}$ a.u.~energy and density convergence criteria were also used to converge the self-consistent field calculation to initialize our TD-(QED-)HF propagation for formaldehyde.}

\section{Results and Discussion}
\label{sec:results_discussion}

\subsection{QED-HF response properties from TD-QED-HF simulations}
\label{sec:results:tdhf}

\begin{figure}[!htbp]
    \centering
    \includegraphics[width=\linewidth]{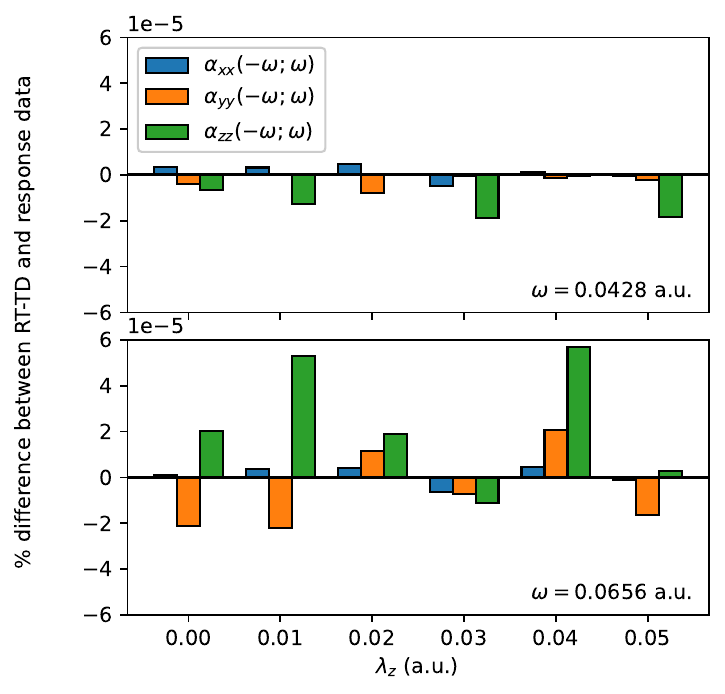}
    \caption{Percent differences between selected Cartesian components of the dynamic polarizability tensor for formaldehyde computed using RT-TD and response QED-HF/aug-cc-pVDZ, for different values of the cavity coupling constant in the $z$ direction ($\lambda_z$). The top and bottom panels correspond to $\omega=0.0428$ a.u. and $\omega=0.0656$ a.u., respectively.}
    \label{fig:comparison_alpha}
\end{figure}

\begin{figure}[!htbp]
    \centering
    \includegraphics[width=\linewidth]{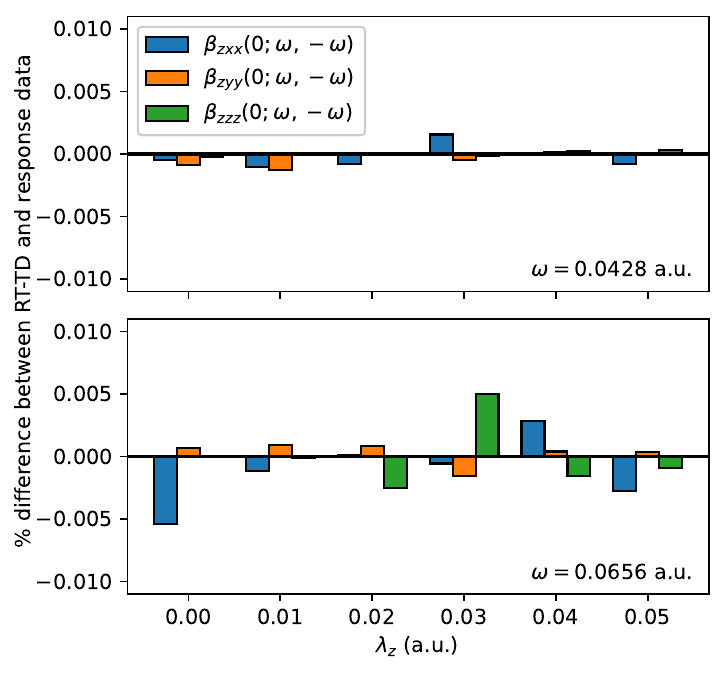}
    \caption{Percent differences between selected Cartesian components of the OR tensor for formaldehyde computed using RT-TD and response QED-HF/aug-cc-pVDZ, for different values of the cavity coupling constant in the $z$ direction ($\lambda_z$). The top and bottom panels correspond to $\omega=0.0428$ a.u. and $\omega=0.0656$ a.u., respectively.}
    \label{fig:comparison_OR}
\end{figure}

\begin{figure}[!htbp]
    \centering
    \includegraphics[width=\linewidth]{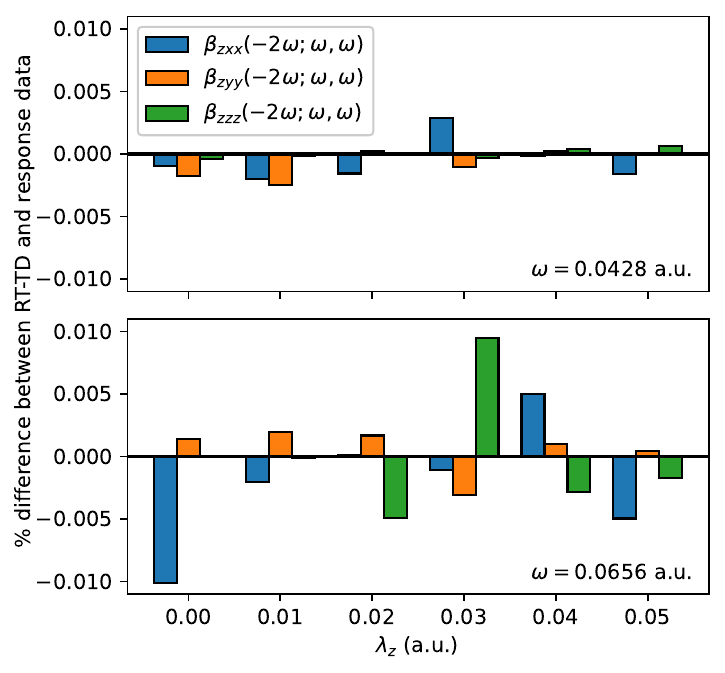}
    \caption{Percent differences between selected Cartesian components of the SHG tensor for formaldehyde computed using RT-TD and response QED-HF/aug-cc-pVDZ, for different values of the cavity coupling constant in the $z$ direction ($\lambda_z$). The top and bottom panels correspond to $\omega=0.0428$ a.u. and $\omega=0.0656$ a.u., respectively.}
    \label{fig:comparison_SHG}
\end{figure}

In this section, we validate our QED-HF response theory implementation numerically against response properties extracted from TD-QED-HF simulations. Figures \ref{fig:comparison_alpha}--\ref{fig:comparison_SHG} contain differences between selected Cartesian components of the dynamic electric dipole  polarizability, OR, and SHG tensors, respectively, obtained from RT-TD- and response-theory-based QED-HF calculations. The raw data, along with values for the same properties at the standard (non-QED) HF level that were obtained from the GAMESS package, are tabulated in Tables S2--S4 of the Supporting Information. The data therein show that response-theory-based QED-HF properties obtained at the cavity-free ($\lambda=0$ a.u.)~limit agree with the corresponding  properties obtained using GAMESS, at both perturbing frequencies ($\omega=0.0428$ a.u.~and 0.0656 a.u.). At $\lambda= 0$ a.u., the components of the dynamic polarizability, OR, and SHG tensors extracted from TD-QED-HF simulations are also in good agreement with their response-theory-based counterparts. For polarizabilities, the agreement is excellent, with deviations of { that are on the order of 0.00001\%} (Fig.~\ref{fig:comparison_alpha}). For the { OR} tensor (Fig.~\ref{fig:comparison_OR}), the agreement is also good, but the deviation grows to { about 0.001\%--0.002\%. However, the $zxx$ component at perturbing frequency $\omega=0.0656$ a.u. shows a larger error of about $-0.005$\%, although it is still insignificant overall}. The SHG results (Fig.~\ref{fig:comparison_SHG}) are qualitatively similar to the OR { ones, but the the overall deviation is larger (e.g., the difference is now about 0.01\% for the $zxx$ component at $\omega=0.0656$ a.u.).} The larger deviations between time-domain- and response-theory-derived components of the first hyperpolarizability tensors can be attributed to the degradation in the quality of the fitting of $\mu_{ABB}^{\text{e},(2)}(t)$ (see Figs.~S1, S2 and Table S1 in the Supporting Information), which could be mitigated by including additional ramping periods. Overall, these data indicate that response properties extracted from time-domain simulations reliably reproduce those from response-theory based calculations at $\lambda = 0$ a.u. They also establish a baseline for expected agreement between our RT-TD- and response-theory-based QED-HF response theory implementations at non-zero coupling strengths.

For non-zero $\lambda$, the data in Fig.~\ref{fig:comparison_alpha} show that the components of the dynamic polarizability tensors extracted from QED-HF response and RT-TD-QED-HF calculations { retain the $\approx0.00001$\% deviation} for all coupling strengths { with several spikes for the $zz$ components that could reach 0.00006\% differences}; this level of agreement is consistent with the cavity-free calculations discussed above and constitutes numerical evidence of the correctness of the QED-HF response theory implementation. The OR and SHG data in Figs.~\ref{fig:comparison_OR} and \ref{fig:comparison_SHG} also exhibit similar patterns at non-zero $\lambda$ as were observed in the cavity-free case. { The deviation in the OR tensors at nonzero coupling constant values remain on the order of 0.001\%--0.002\%, with a few spikes that can be as large as 0.01\%.} Furthermore, the SHG results are qualitatively similar to the OR ones; { again, the deviations in the SHG values tend to be larger than in OR, and the spikes in the percent difference appear on the same combinations of $\lambda_z$ value and Cartesian components.} As in the zero-coupling limit, the larger variations in the first hyperpolarizabilities can be traced to the { slight} reduction in fit quality for $\mu_{ABB}^{\text{e},(2)}(t)$ compared to that of $\mu_{AA}^{\text{e},(1)}(t)$, as seen in the $R^2$ values reported in Table S1 in the Supporting Information. Overall, the differences between time-domain- and response-theory-based QED-HF properties obtained at zero and non-zero coupling strengths are comparable, and, importantly, we observe no clear trend suggesting that increasing coupling strength leads to increasing deviations { in a systematic manner}. Put together, these observations provide numerical evidence of the consistency and correctness of our implementations of TD-QED-HF and the QED-HF response equations.

{ Before moving on, let us comment on the parameter choices of our RT-TD propagations. Due to the nonlinear EOMs in the time evolution of (QED-)HF density matrices, the RK4 integrator may not be the best choice for this problem. For example, RT-TD-DFT calculations in Ref.~\citenum{Li13_064104} used the modified midpoint unitary transformation approach,\cite{Schlegel05_233,Tully07_134307} with the argument that Runge--Kutta-type integrators require small time steps to preserve the idempotency of the density matrix.\cite{Schlegel05_233} Furthermore, although they were employed for correlated approaches, the details of time integrators in Refs.~\citenum{Pedersen23_154102,Crawford25_1908} may also be of interest to the reader. Nevertheless, as shown by the excellent agreement shown in Figs.~\ref{fig:comparison_alpha}--\ref{fig:comparison_SHG}, RK4 can still be acceptable, provided that suitable parameters are employed (\emph{cf.}~Figs.~S14 and S15 in the Supporting Information). In particular, one still needs to employ rather long ramping period, and the field strength cannot be too large (which results in the appearance of rapid oscillations) or too small (in which case the numerical noise washes out the oscillations of interest).}

\subsection{Cavity-induced changes to dynamic response properties}

In this section, we explore how cavity interactions modulate frequency-dependent response properties in molecular systems computed using QED-HF response theory. Simulation results are summarized in Figs.~\ref{fig:alpha_combined}--\ref{fig:SHG_combined}, where we focus on the $\alpha_{AA}$ and $\beta_{zAA}$ components of the (hyper)polarizability tensors with the cavity polarized along the $A$ axis, for $A\in\{x,y,z\}$. The response properties are reported as a percent change with respect to the value at the cavity-free ($\lambda=0$ a.u.) limit. Additional data involving other Cartesian components of the tensors and different choices for the cavity mode polarization axis are provided in the Supporting Information.

\subsubsection{Dynamic polarizability}

\begin{figure*}[!htbp]
    \centering
    \includegraphics[width=\linewidth]{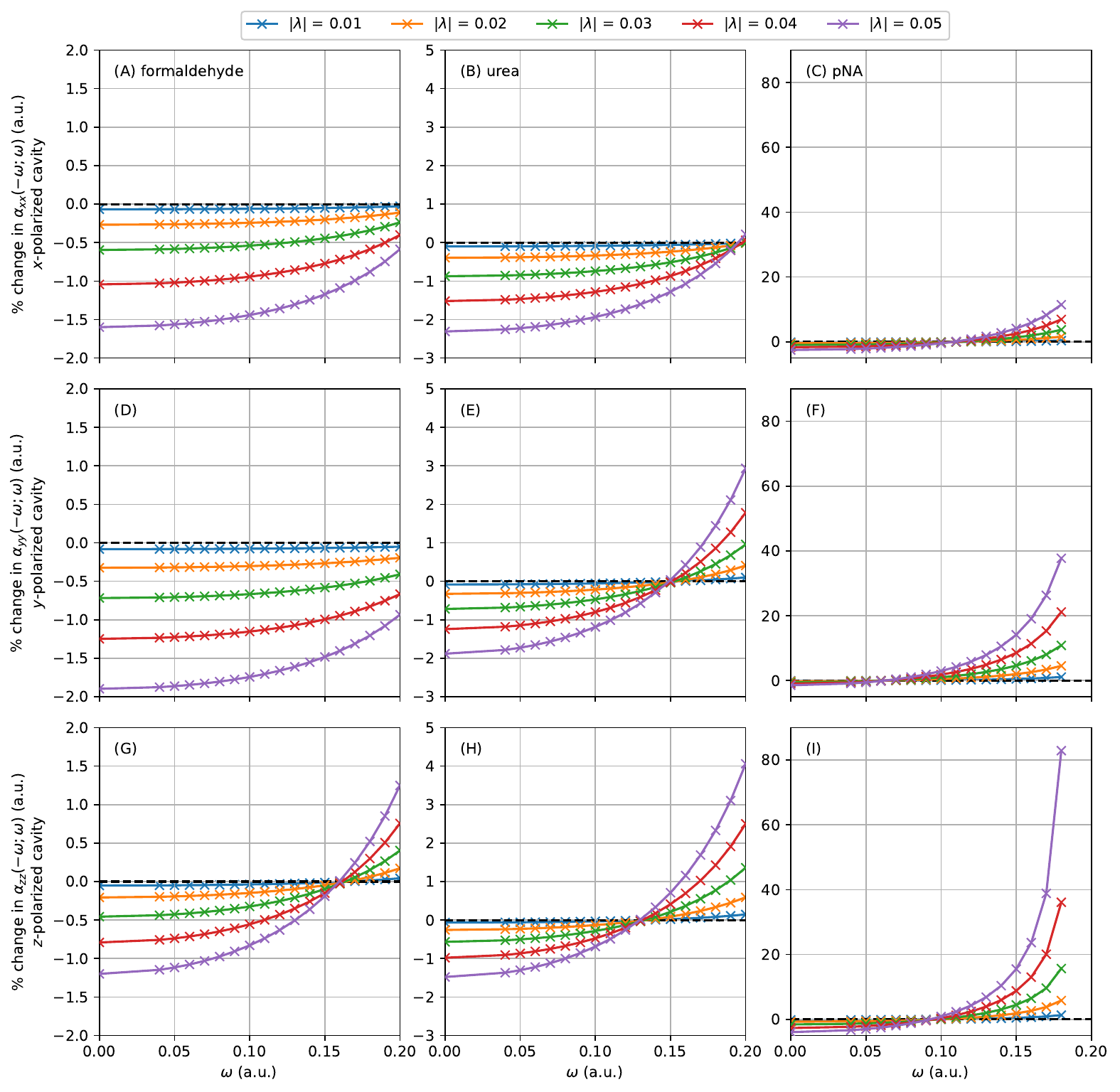}
    \caption{Cavity-induced changes in the dynamic polarizabilities of formaldehyde (left column), urea (middle column), and pNA (right column), computed using (QED-)HF/d-aug-cc-pVTZ, as functions of perturbing frequency $\omega$. The top, middle, and bottom rows correspond to the $xx$, $yy$, and $zz$ Cartesian components of $\alpha(-\omega;\omega)$, as well as cavity polarization along the $x$, $y$, and $z$ axes. Each data point is reported as percent difference relative to the $\lambda=0$ a.u.~value.}
    \label{fig:alpha_combined}
\end{figure*}

We begin by examining the influence of the cavity on the dynamic polarizability tensors for formaldehyde, urea, and pNA. Figure \ref{fig:alpha_combined} depicts the percent changes in $\alpha_{xx}(-\omega;\omega)$, $\alpha_{yy}(-\omega;\omega)$, and $\alpha_{zz}(-\omega; \omega)$ (top, middle, and bottom panels, respectively) as functions of the perturbing frequency, $\omega$. In the static limit ($\omega=0$ a.u.), the polarizabilities are suppressed with increasing $\lambda$ value, for all molecules and Cartesian components of the polarizability tensor. This result is consistent with the trend reported for the isotropically averaged static polarizability computed for pNA in our earlier work.~\cite{DePrince25_8024} The magnitude of the largest cavity-induced changes to the static polarizabilities correlate positively with the size of the molecule, ranging from 1.2\%--1.9\% for formaldehyde to 1.5\%--2.3\% for urea to 1.5\%--4\% for pNA.

The diagonal components of the dynamic polarizability tensor increase in magnitude with increasing perturbing frequency (see the raw data in the Supporting Information). 
{This behavior is consistent with the familiar sum-over-states expression for the polarizability, {\em i.e.}, the polarizability increases as the perturbing frequency approaches an excitation energy from the left.} Let us focus first on pNA, where the lowest-energy poles in the $\alpha_{xx}$, $\alpha_{yy}$, and $\alpha_{zz}$ response functions at $\lambda = 0$ a.u.~occur at 0.210, 0.201, and 0.192 a.u., respectively.  For non-zero coupling strengths, we observe a rapidly increasing percent change in the polarizability for different $\lambda$ values in panels (C), (F), and (I) of Fig.~\ref{fig:alpha_combined} near these poles, the positions of which are slightly perturbed by the cavity (see Fig.~S12 in the Supporting Information), despite the fact that the cavity is tuned to { a} higher-energy bright $A_1$-symmetry state (see Table \ref{tab:vee_and_os}). At a coupling strength of $\lambda = 0.05$ a.u., the diagonal components of the polarizability tensors are enhanced by up to 11\% ($\alpha_{xx}$), 38\% ($\alpha_{yy}$), and 83\% ($\alpha_{zz}$) (at $\omega = 0.18$ a.u.). The massive enhancement of the $zz$ component of the polarizability tensor can be explained, at least in part, by the fact that the lowest-energy pole in the response function shifts to lower energies with increasing coupling strengths (see Fig.~S12 in the Supporting Information). As a result, the polarizability increases more rapidly at lower perturbing frequencies than in the cavity-free case, leading to the large enhancement seen in panel (I) of Fig.~\ref{fig:alpha_combined}. However, the shift in the pole location cannot completely explain the data here, particularly for the $xx$ and $yy$ components of the polarizability, where the cavity effect on the position of the pole is much smaller (Fig.~S12). For formaldehyde and urea, the lowest-energy poles at $\lambda = 0$ a.u.~are closer to 0.3 a.u., so we do not observe such a dramatic increase in the values of the diagonal elements of the polarizability tensor. As was the case for pNA, the percent change in the polarizabilities becomes less negative / more positive as the perturbing frequency is increased, although the effect is not as dramatic as for pNA, at least in the frequency range considered in Fig.~\ref{fig:alpha_combined}. 

As mentioned, the diagonal elements of the polarizability tensors for all molecules increase with increasing perturbing frequency, at all coupling strengths. At non-zero $\lambda$, all molecules also show a percent change in $\alpha_{AA}(-\omega;\omega)$ that becomes less negative / more positive with increasing $\omega$, when the cavity mode is polarized along the $A$ direction. Again, we note that data corresponding to other polarization choices are provided in the Supporting Information; Figs.~S3, S6, and S9 shows a different behavior where $\alpha_{AA}(-\omega;\omega)$ is suppressed at all perturbing frequencies when the cavity mode is polarized along a direction orthogonal to $A$ (with the exception of the $\alpha_{zz}$ component for pNA, with a $y$-polarized cavity mode). Regardless, focusing on the data on Fig.~\ref{fig:alpha_combined}, we see that, with the exception of the $\alpha_{xx}$ and $\alpha_{yy}$ response functions for formaldehyde, which are suppressed by the cavity over the range of $\omega$ values considered in Fig.~\ref{fig:alpha_combined}, all other panels of this figure show that there exists a particular perturbing frequency at which the percent change in the polarizability passes through zero, which indicates a shift from cavity-induced suppression of the property in the static limit to enhancements at higher perturbing frequencies. This crossover frequency is different for each molecule and Cartesian component of the polarizability tensor, but, interestingly, it appears to be independent of the cavity coupling strength. In other words, this special perturbing frequency represents an $\omega$ value at which there is no cavity effect on the response property, regardless of the magnitude of $\lambda$. Note that the existence of this  feature is not a resonance effect; recall that, for each molecule, the cavity mode frequency is tuned to the same bright electronic excitation, regardless of the polarization of the cavity mode. In cases where the cavity mode cannot couple to the electronic state in question by symmetry ({\em i.e.}, for $x$- and $y$-polarized cavity modes), there is no polariton formation. We also stress that the existence of this feature cannot simply be attributed to cavity-induced shifts in the location of the lowest-energy poles in the response function. We noted above that the sharp enhancement of the $\alpha_{zz}$ function in the case of pNA could be traced to a cavity-induced shift of the pole to a lower perturbing frequency, but the situation is more complex, in general. For urea, for example, the cavity has the effect that it shifts the lowest-energy poles in the response function to {\em higher} perturbing frequencies (see Fig.~S12 of the Supporting Information). If this shift was the only effect in play, then the diagonal components of the polarizability tensor for urea would be suppressed, not enhanced, at higher perturbing frequencies, and there would be no crossover perturbing frequency in panels (B), (E), and (H) of Fig.~\ref{fig:alpha_combined}.  

{We can show mathematically that the crossover frequency observed in Fig.~\ref{fig:alpha_combined} is coupling-strenght independent. Recall from Eq.~\ref{EQN:POLARIZABILITY_CROSSING} in Sec.~\ref{SEC:RESPONSE_THEORY} that the polarizability for the cavity-coupled system can be decomposed into a sum of components that include either DSE or BLC effects. The cavity-induced change to the polarizability can be decomposed in the same way. Consider the following expression for the difference in the polarizability for the free and cavity-embedded molecular systems
\begin{align}
    \Delta \alpha_{DC}(\omega) &= \alpha_{DC}^\text{cav}(-\omega;\omega) - \alpha_{DC}^\text{free} (-\omega;\omega) \nonumber \\
    & = [\alpha^0_{DC}(-\omega;\omega) - \alpha_{DC}^\text{free}(-\omega;\omega)] \nonumber \\
    &+ \frac{\tau(\omega)\sigma_C^0(\omega)\alpha_{D,\lambda}^0(\omega)}{1-\tau(\omega)\gamma(\omega)} \nonumber \\
    &= \delta\alpha_{DC}^\text{DSE}(\omega) + \delta\alpha_{DC}^\text{BLC}(\omega)
\end{align}
where $\alpha_{DC}^\text{cav}(-\omega;\omega)$ and $\alpha_{DC}^\text{free}(-\omega;\omega)$ are the polarizabilities of the molecule inside and outside of a cavity, respectively, and we have introduced the symbols 
\begin{align}
    \delta\alpha_{DC}^\text{DSE}(\omega) = \alpha^0_{DC}(-\omega;\omega) - \alpha_{DC}^\text{free}(-\omega;\omega)
\end{align}
and
\begin{align}
    \delta\alpha_{DC}^\text{BLC}(\omega) = \frac{\tau(\omega)\sigma_C^0(\omega)\alpha_{D,\lambda}^0(\omega)}{1-\tau(\omega)\gamma(\omega)}
\end{align}
to refer to the DSE- and BLC-derived contributions to the polarizability change. The crossover frequency, $\omega^*$, is the frequency at which the cavity induces no change in the polarizability and is defined by
\begin{align}
\delta\alpha_{DC}^\text{DSE}(\omega^*) = - \delta\alpha_{DC}^\text{BLC}(\omega^*)
\end{align}

We can show that this frequency condition is coupling-strength independent by considering the simplified case involving only the diagonal $zz$ element of the polarizability tensor for a molecule interacting with a $z$-polarized cavity mode (\emph{i.e.}, with ${\bm \lambda} = [0, 0, \lambda]$). At small $\lambda$, we can expand $\delta\alpha^\text{DSE}_{zz}(\omega^*)$ in a Taylor series in $\lambda$, which, when truncated at second order is
\begin{align}
\delta\alpha^\text{DSE}_{zz}(\omega^*) \approx  \frac{1}{2}\bigg ( \frac{\partial^2 \alpha_{zz}^0(\omega^*)}{\partial \lambda^2}\bigg ) \bigg |_{\lambda = 0}\lambda^2 + \mathcal{O}(\lambda^4)
\end{align}
where the series is even in $\lambda$ because the DSE term is quadratic in $\lambda$ and, by definition, the BLC term does not contribute to $\alpha_{zz}^0(\omega^*)$. Now, expanding  
$\delta\alpha_{zz}^\text{BLC}(\omega^*)$ in the same way gives
\begin{align}
     \delta\alpha_{zz}^\text{BLC}(\omega^*) &\approx \frac{1}{2} \lambda^2\bigg ( \frac{\partial^2}{\partial \lambda^2} \frac{\tau(\omega^*)\sigma_z^0(\omega^*)\alpha_{z,\lambda}^0(\omega^*)}{1-\tau(\omega^*)\gamma(\omega^*)}\bigg ) \bigg |_{\lambda = 0} + \mathcal{O}(\lambda^4) \nonumber \\
    &=\frac{1}{2} \lambda^2\bigg ( \frac{\partial^2}{\partial \lambda^2}\frac{\lambda^2\tau(\omega^*) [\alpha_{zz}^0(\omega^*)]^2}{1-\lambda^2 \alpha_{zz}^0(\omega^*)\tau(\omega^*)}\bigg ) \bigg |_{\lambda = 0} + \mathcal{O}(\lambda^4) \nonumber \\
    &= \lambda^2\tau(\omega^*) [\alpha_{zz}^\text{free}(-\omega^*;\omega^*)]^2+ \mathcal{O}(\lambda^4)
\end{align}
where, again, the odd powers in $\lambda$ in the series vanish.
To leading order in $\lambda$, the crossover frequency condition becomes
\begin{align}
    \frac{1}{2}\bigg ( \frac{\partial^2 \alpha_{zz}^0(\omega^*)}{\partial \lambda^2}\bigg ) \bigg |_{\lambda = 0}\lambda^2 = -\lambda^2\tau(\omega^*) [\alpha_{zz}^\text{free}(-\omega^*;\omega^*)]^2
\end{align}
or 
\begin{align}
    \frac{1}{2}\bigg ( \frac{\partial^2 \alpha_{zz}^0(\omega^*)}{\partial \lambda^2}\bigg ) \bigg |_{\lambda = 0} = -\tau(\omega^*) [\alpha_{zz}^\text{free}(-\omega^*;\omega^*)]^2
\end{align}
Remarkably, this condition does not depend explicitly on the coupling strength, which is consistent with our observations in Fig.~\ref{fig:alpha_combined}.
}

\subsubsection{First hyperpolarizabilities: OR and SHG}

\begin{figure*}[!htbp]
    \centering
    \includegraphics[width=\linewidth]{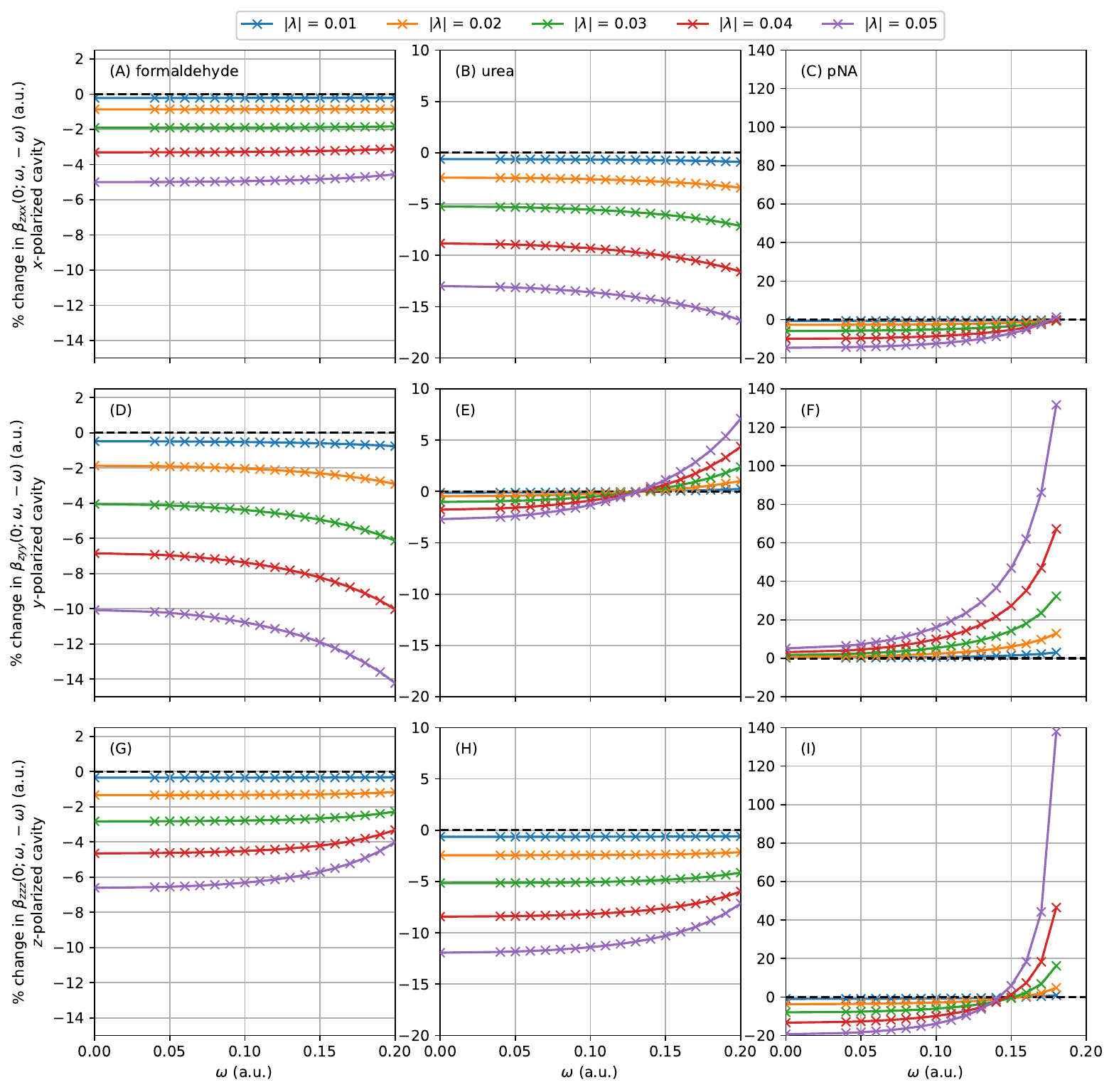}
    \caption{Cavity-induced changes in the OR tensor of formaldehyde (left column), urea (middle column), and pNA (right column), computed using (QED-)HF/d-aug-cc-pVTZ, as functions of perturbing frequency $\omega$. The top, middle, and bottom rows correspond to the $zxx$, $zyy$, and $zzz$ Cartesian components of $\beta(0;\omega,-\omega)$, as well as cavity polarization along the $x$, $y$, and $z$ axes. Each data point is reported as percent difference relative to the $\lambda=0$ a.u.~value.}
    \label{fig:OR_combined}
\end{figure*}

\begin{figure*}[!htbp]
    \centering
    \includegraphics[width=\linewidth]{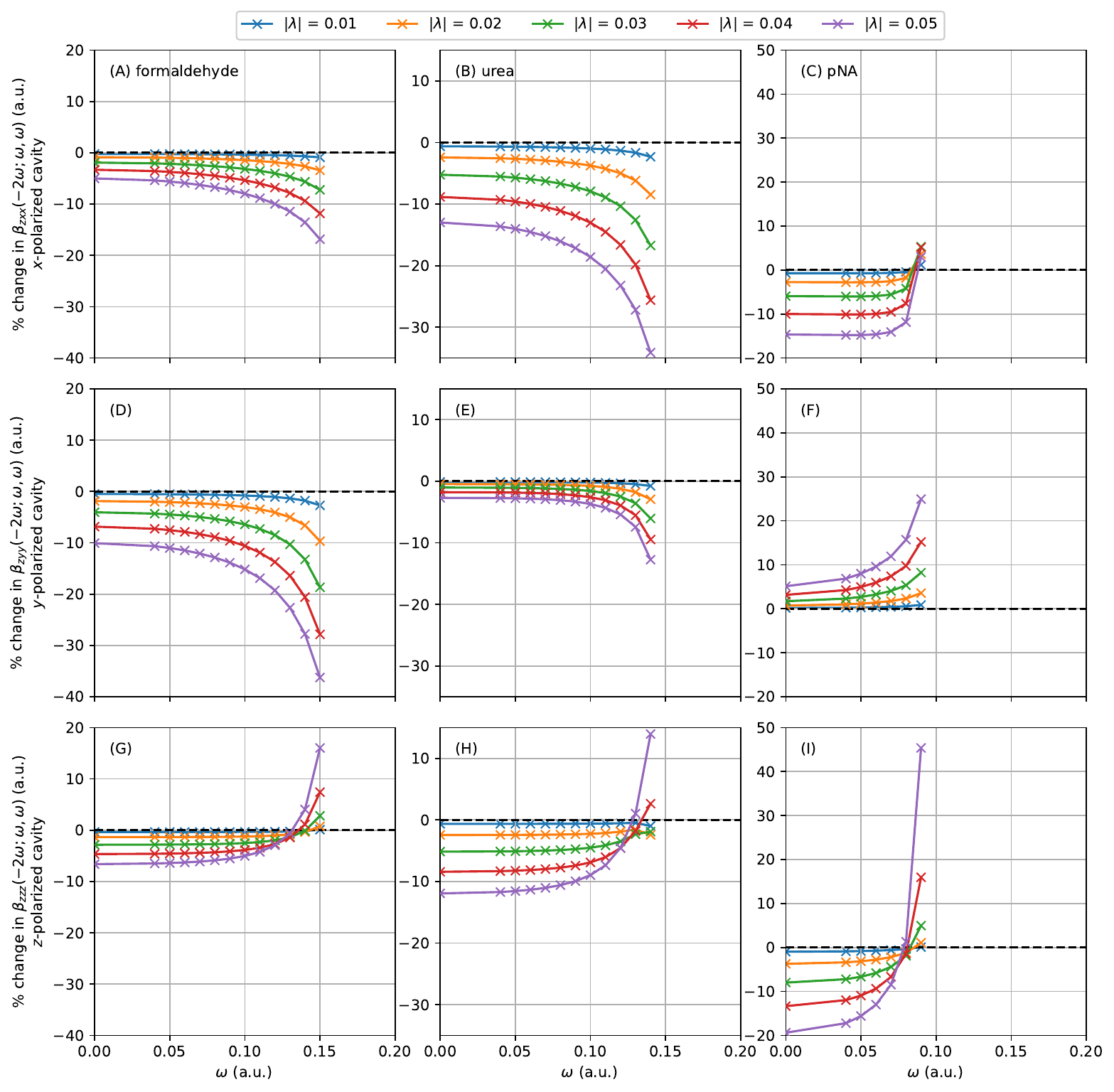}
    \caption{Cavity-induced changes in the SHG tensor of formaldehyde (left column), urea (middle column), and pNA (right column), computed using (QED-)HF/d-aug-cc-pVTZ, as functions of perturbing frequency $\omega$. The top, middle, and bottom rows correspond to the $zxx$, $zyy$, and $zzz$ Cartesian components of $\beta(-2\omega;\omega,\omega)$, as well as cavity polarization along the $x$, $y$, and $z$ axes. Each data point is reported as percent difference relative to the $\lambda=0$ a.u.~value.}
    \label{fig:SHG_combined}
\end{figure*}

Figures~\ref{fig:OR_combined} and \ref{fig:SHG_combined} 
report select Cartesian components the $\beta(0; \omega,-\omega)$ and $\beta(-2\omega; \omega, \omega)$ tensors, respectively, for formaldehyde, urea, and pNA. As mentioned earlier, we focus on the $\beta_{zAA}$ elements, for $A\in\{x,y,z\}$, where the cavity mode is polarized along the $A$ direction. Additional data involving other Cartesian components of the tensors and different choices for the cavity mode polarization axis are provided in the Supporting Information. Let us first consider the static case, where the OR and SHG tensors recover the same limit. At $\omega = 0$ a.u., the influence of the cavity on the first hyperpolarizabilities differs from that on the static polarizabilities in several key ways. First, the data depicted in Fig.~\ref{fig:alpha_combined} show that the magnitudes of the diagonal components of the static polarizability tensor decrease with increasing $\lambda$, for all molecules. Here, we find similar suppression of the static $\beta_{zAA}$ values for all $A$ for formaldehyde and urea, as well as the $\beta_{zxx}$ and $\beta_{zzz}$ components for pNA, but the $\beta_{zyy}$ component for pNA is  enhanced by the cavity at the static limit. We noted a similar orientation-dependent cavity effect for the isotropic static hyperpolarizability in Ref.~\citenum{DePrince25_8024}.  Second, the magnitude of the cavity effect is significantly larger for the static hyperpolarizability than for the static polarizability. Figures~\ref{fig:OR_combined} and \ref{fig:SHG_combined} show that, for each molecule, at least one of the $\beta_{zAA}$ components shows cavity induced suppression or enhancement that reaches or exceeds 10\% in magnitude at the static limit. This increased sensitivity to the cavity was also observed for the isotropically averaged static hyperpolarizability versus polarizability in Ref.~\citenum{DePrince25_8024}.

Now, let us consider the frequency dependence of the hyperpolarizability tensors. First, the magnitudes of the components of the hyperpolarizability tensors depicted in Figs.~\ref{fig:OR_combined} and \ref{fig:SHG_combined} increase as we approach poles in the response functions, as was also the case for the diagonal components of the polarizability tensors. For OR, the poles are in the same locations as for the polarizability because both properties require solving for response amplitudes at the perturbing frequency, $\omega$. For SHG, we have additional poles at lower perturbing frequencies because this response function depends on response amplitudes determined at both $\omega$ and $-2\omega$. These additional poles make the numerical problem of solving for the amplitudes more challenging, which is why there are more missing data points in Fig.~\ref{fig:SHG_combined} than Fig.~\ref{fig:OR_combined}. 

The frequency dependence of the cavity-induced changes to the first hyperpolarizability tensors is complex. The OR data depicted in Fig.~\ref{fig:OR_combined} show both increased suppression in panels (B) and (D) and decreased suppression and sometimes enhancement in all other panels. The SHG data in Fig.~\ref{fig:SHG_combined} tell a similar story, with increased cavity-induced suppression in panels (A), (B), (D), and (E), and decreased suppression / enhancement elsewhere. As in the static case, the dynamic hyperpolarizabilities are far more sensitive to cavity effects than the static polarizabilities. For pNA, in particular, we observe nearly 140\% increases in $\beta_{zyy}(0; \omega, -\omega)$ and $\beta_{zzz}(0; \omega, -\omega)$ values as we approach the first pole in the response function [panels (F) and (I) of Fig.~\ref{fig:OR_combined}], and a nearly 50\% increase in $\beta_{zzz}(-2\omega; \omega, \omega)$ [panel (I) of Fig.~\ref{fig:SHG_combined}]. In some cases, the sharp enhancement / suppression near poles in the response functions can be partially rationalized by cavity-induced shifts to the locations of the poles. For example, in the case of OR for urea, the lowest-energy pole in $\beta_{zAA}(0; \omega, -\omega)$ shifts to higher energies upon coupling to the cavity (see Fig.~S12 of the Supporting Information). This shift could explain the enhanced suppression of the $\beta_{zxx}(0; \omega, -\omega)$ and $\beta_{zyy}(0; \omega, -\omega)$ components, but not the enhancement of the $\beta_{zzz}(0; \omega, -\omega)$ component. As such, the cavity-induced changes to the locations of the poles cannot be the only effect at play. 

As we saw in the diagonal components of the dynamic polarizability tensors, we find several crossover perturbing frequencies at which certain components of the hyperpolarizabilty tensors transition from being suppressed by the cavity to being enhanced; Panels (C), (E), and (I) of Fig.~\ref{fig:OR_combined} show this feature for OR, and for SHG, the feature appears in panels (C), (G), (H), (I) of Fig.~\ref{fig:SHG_combined}.  Unlike in Fig.~\ref{fig:alpha_combined}, we also observe situations that preclude the existence of these special frequencies, {\em e.g.}, when the static limit shows suppression by the cavity, with additional suppression at higher perturbing frequencies [panels (B) and (D) of Fig.~\ref{fig:OR_combined}; panels (A), (B), (D), and (E) of Fig.~\ref{fig:SHG_combined}], or, conversely, when the static limit shows enhancement, and additional enhancement occurs for higher perturbing frequencies [panel (F) in Figs.~\ref{fig:OR_combined} and \ref{fig:SHG_combined}]. Again, cavity-induced changes to the locations of the poles in the response functions may contribute to the existence of these special perturbing frequencies, but these shifts { alone} cannot explain their appearance in all cases. 

\section{Conclusions}
\label{sec:conclusions}

We have implemented linear and quadratic response functions for cavity QED-HF wave functions in order to evaluate cavity-induced changes to dynamic electric dipole polarizability and first hyperpolarizability tensors. 
{To the best of our knowledge, this is the first such implementation of quadratic response functions within a cavity QED formalism.} The polarizability, OR, and SHG tensors generated via response theory were numerically verified against an independent implementation where the response properties were {determined via}
the TD finite-field approach described in Refs.~\citenum{Li13_064104,Pedersen23_154102,Crawford25_1908}{, which was extended here, for the first time, to the cavity QED case}. We have demonstrated both suppression and enhancement of response properties near poles in the response functions, with the hyperpolarizability tensors being particularly sensitive to cavity effects. Moreover, simulations involving multiple cavity coupling constants and a range of perturbing frequencies revealed special perturbing frequencies at which components of the response tensors show negligible cavity-induced changes, regardless of the magnitude of the cavity coupling strength. The existence of these crossover frequencies suggest that cavity-induced suppression or enhancement of a response property could be controlled by the perturbing frequency. Notably, the specific numerical value of this special perturbing frequency does not obviously relate to the cavity mode frequency itself, nor does it depend on resonant interactions or polariton formation. {A perturbative analysis of the crossover frequency condition confirms that the crossover frequency is coupling-strength independent, at least at the lowest order in $\lambda$.}

{ Lastly, we note that prior work suggests that cavity-induced changes in electronic properties may be artificially amplified by methods that treat electron-photon interactions at the mean-field level.\cite{DePrince21_094112,DePrince23_5264,DePrince24_064109} As such, it will be interesting to explore how electron-electron and electron-photon correlation effects impact the degree to which the cavity alters molecular response properties, as well as the existence and character of the observed crossover frequency.}

\vspace{0.5cm}

{\bf Supporting Information} { A derivation of the EOM for the electronic density matrix; the} optimized geometries for formaldehyde, urea, { pNA, and formate}; fitting data for $\mu_{AA}^{\text{e},(1)}(t)$ and $\mu_{ABB}^{\text{e},(2)}(t)$ for formaldehyde in the aug-cc-pVDZ basis; additional dynamic polarizability, OR, and SHG data with different cavity polarization choices; the lowest poles in the formaldehyde, urea, and pNA molecules within the d-aug-cc-pVTZ basis set; { dynamic polarizabilities for formate anion with the center of mass at origin and shifted by 10 \AA; parameter testings for the TD-(QED-)HF propagations;} and a zipped archive of three CSV files containing the computed $\alpha(-\omega;\omega)$, $\beta(0;\omega,-\omega)$, and $\beta(-2\omega;\omega,\omega)$ for formaldehyde, urea, and pNA, respectively.

\vspace{0.5cm}

\ifacs
    \noindent {\bf ACKNOWLEDGMENTS:} \\
\else
    \begin{acknowledgments}
\fi
We thank the National Science Foundation -- Research Experiences for Undergraduates Sites program. This material is based upon work supported by the NSF under Grants CHE-2150301 and CHE-2533603. SHY acknowledges support from the Florida State University Quantum Initiative. The computing for this project was performed on the high performance computing (HPC) cluster at the Florida State University Research Computing Center. { The authors acknowledge the use of AI tools (Claude) to assist in the analysis of the origin of the crossover frequency discussed in Sec.~\ref{sec:results_discussion}.}\\
\ifacs
\else
    \end{acknowledgments}
\fi

\noindent {\bf AVAILABILITY}\\

The numerical data and molecular geometries underlying this study are available in the Supporting Information. The C++ code used to generate the response-theory-based QED-HF data is available at https://github.com/edeprince3/hilbert. Example input files for running QED-HF dynamic response calculations are provided in the tests directory therein. The Python code to generate response properties from TD-QED-HF simulations is available at https://github.com/stephenyuwono/dumb-polaritonic-hf.

\bibliography{Journal_Short_Name,main}

\end{document}